\documentclass[12pt,a4paper]{extarticle}
\usepackage[english]{babel}
\usepackage{hyperref}
\usepackage[numbers]{natbib}
\setcitestyle{numbers,open={[},close={]}}

\makeatletter
\def\@rmpt#1.{#1}
\def\put@pt#1#2.#3{\ifx#3\empty\@rmpt#2#1\else #2.\kern-.25em\relax#1\@rmpt#3\fi}
\def\m#1{\ensuremath{\put@pt{^m}#1.\empty}}
\def\arcdeg#1{\ensuremath{\put@pt{^\circ}#1.\empty}}
\def\arcmin#1{\ensuremath{\put@pt{'}#1.\empty}}
\def\arcsec#1{\ensuremath{\put@pt{''}#1.\empty}}

\makeatother

\def\saoname{Special Astrophysical Observatory  RAS,  Nizhnii Arkhyz, 369167 Russia}

\begin{document}


\title{\bf Stellar Spectroscopy Technique on Small- and Intermediate-Diameter Telescopes}

\author{V.E.~Panchuk,  V.G.~Klochkova \& E.V.~Emelianov\\
     \centerline{\small\href{mailto:panchuk@ya.ru}{panchuk@yandex.ru}}\\
 \centerline{\small\saoname}}
\date{}

\maketitle

\abstract{
We briefly present the history of technical solutions aimed at improving the
efficiency of spectroscopy on small- and moderate-diameter telescopes. We assess
the current state of spectroscopy techniques and some of the perspectives.
}

\section{Introduction}
The interest in small- and moderate-diameter telescopes is due to the
development of methods for the use of novel detectors. Thus the results of the
first revolution of the technique of registration of weak signals (mind
XX-century) served as the basis for the program of the 1956
symposium~\cite{bradshaw1958present} (the Russian edition was also
published~\cite{scheglov1960present}, which was supplemented with papers
considered to be promising). The next time the subject came to the focus at the
level of an IAU symposium was in 1986 \cite{hearnshaw1985instrumentation}, when
the results of the use of single-channel detectors were summarized and the
prospects of the application of multichannel solid-state detectors already
became apparent. Whereas in the mid-XX century instruments with diameters of 0.3
to 1-m were viewed as moderate-diameter telescopes, 30 years later telescopes
with diameters up to 1 m were already classified as small instruments. A brief
review of the types of equipment used on small- and moderate-diameter telescopes
was presented by Panchuk et.~al.~\cite{panchuk2004prepsao195}, and in 2015 the international
meeting ``The Present and Future of Small and Medium Size
Telescopes''~\footnote{\protect\url{https://www.sao.ru/hq/lon/ConfSite/program-ru.html}} was held
at the Special Astrophysical Observatory of the Russian Academy of Sciences.

We restrict the ensuing review of stellar spectroscopy techniques used on small-
and moderate-diameter telescopes to the \mbox{$D\sim[0.25; 1.22]$~m} diameter
interval. In exceptional cases when discussing new methodological achievements
we mention studies carried out on  \mbox{$D\sim1.5$~m} telescopes. Unlike our
review~\cite{panchuk2004prepsao195}, we do not mention photometry techniques used on
small-diameter telescopes. Such studies have since long become a separate field.

First, we provide some general considerations concerning small- and
moderate-diameter  telescopes, which we partly pointed out in our earlier
paper~\cite{panchuk2004prepsao195}. The advantages and specificities of the
operating of small- and moderate-diameter telescopes can be conveniently
classified into technical, financial, organizational, scientific, and
psychological types.

\newpage
\begin{list}{}{
\setlength\leftmargin{2mm} \setlength\topsep{2mm}
\setlength\parsep{0mm} \setlength\itemsep{2mm} }
 \item \underline{Technical:}
 \begin{list}{$\bullet$}{
\setlength\leftmargin{4mm} \setlength\topsep{2mm}
\setlength\parsep{0mm} \setlength\itemsep{2mm} }
 \item Practically all large telescopes are multiprogram instruments, which, unlike small telescopes,
are difficult to optimize for addressing separate tasks. The DAO
reflector ($D=1.22$~m) used for studies involving photographic
spectroscopy in the coud\'e focus is a classical example of this
type: the spectrograph of this telescope~\cite{richardson1968}
outperformed the coud\'e spectrograph of the Hale telescope
($D=5$~m) in terms of limiting
magnitude~\cite{bowen1952thespectrographic}.

\item Because of the losses at the entrance slit, the advantage in
the limiting magnitude in the case of high-resolution spectroscopy
performed on intermediate-diameter and large telescopes is
proportional to the first power of the mirror diameter rather than
to the squared diameter as in the case of small telescopes.

\item Small telescopes are easier to automate.
\end{list}

 \item \underline{Financial:}
\begin{list}{$\bullet$}{
\setlength\leftmargin{4mm} \setlength\topsep{2mm}
\setlength\parsep{0mm} \setlength\itemsep{2mm} }
 \item If a small telescope is equipped only with one kind of instruments,
 it is guaranteed to be operated at full capacity.
The per-observation cost, which is determined not only by the
operating costs but also by the cost of   auxiliary  instrumentation,
is thereby reduced.

 \item If a small telescope is narrowly specialized (one type of equipment
 for dark nights and one type of equipment for
darkless nights), then the expenses for switching the instruments
are small.

 \item Overall technical maintenance of a small telescope is less expensive.

 \item  The current dependence of the cost of a telescope on the
 diameter of its primary mirror shows a sharp increase at
1.2~m \cite{swift2015jast}. Whereas the cost of ``amateur-class''
telescopes is proportional to its diameter, the cost of
professional telescopes (starting from $D\sim1.2$~m) is
proportional to the area of the mirror.
\end{list}

 \item \underline{Organizational:}
\begin{list}{$\bullet$}{
\setlength\leftmargin{4mm} \setlength\topsep{2mm}
\setlength\parsep{0mm} \setlength\itemsep{2mm} }
 \item With small telescopes it is easier to acquire time for repeating observations, which can be necessary for verifying some of the
results.

 \item Equipment can remain longer on a small telescope during the checkout and preparation stage equipment than on a larger-diameter
telescope.

 \item Observations on small telescopes are easier to organize because one observer is enough in most of the cases.

 \item Equipment breakdown on a small telescope is not viewed as a permanent loss of time, and the replacement of equipment is less
formalized in most of the cases.
\end{list}
 \item \underline{Scientific:}
\begin{list}{$\bullet$}{
\setlength\leftmargin{4mm} \setlength\topsep{2mm}
\setlength\parsep{0mm} \setlength\itemsep{2mm} }
 \item Cloudy weather may simultaneously  ``switch off'' and entire
 group of large telescopes located at the same site, whereas
small telescopes better distributed across the Earth surface, have
a better chance to detect a unique event (this does not apply to
extremely faint and short-lived phenomena).

 \item A number of research tasks requires uninterrupted monitoring
 the object with telescopes located at different longitudes (such
monitoring is practically impossible to organize with large
multiprogram telescopes).

 \item Small telescopes offer more time for continued studies, which is important for studying spectral
variability or when carrying out mass spectroscopic surveys. The
efficiency of these instruments also increases as some of the
telescopes mentioned above become multiprogram instruments.

 \item There is a view  \cite{warner1986instrumentation} that the time for fulfilling a single
 observational project on a 1-m
telescope exceeds the corresponding time for a \mbox{4-m}
telescope, whereas one large telescope is equivalent to four
half-diameter telescopes in terms of scientific efficiency. There
are also other scientometric estimates \cite{abt2012aj}.

 \item A number of new phenomena were discovered just with small telescopes.
 The expansion of the Crab nebula was measured
on a $D=0.91$~m telescope \cite{mayall1937pasp49101}. The
same instrument was used in observations that resulted in the
discovery of the rotation of the  M\,31 nebula
\cite{babcock1939lick}. Circular polarization of white dwarfs,
which was first measured with a \mbox{$D=0.61$~m} telescope
\cite{kemp1970apj}, was later studied with larger instruments
\cite{angel1971apj}. Mass photoelectric radial-velocity
measurements, which, in particular, became the basis for compiling
programs for Doppler-based searches for low-mass companions, were
carried out with meter-class telescopes \cite{griffin1967apj,
fletcher1982pasp, mayor1986stellar}. The first photometric studies
of nonradial pulsations were performed on \mbox {$D=[0.6; 0.9]$}~m
telescopes \cite{kurtz1982mnras}. The studies of optical effects
accompanying short-wavelength flares are performed on even smaller
telescopes. Some of the effects discovered with large telescopes
were then studied in detail with small instruments. Linear
polarization, which was discovered in four stars in observations
carried out on a $D=2.08$~m telescope \cite{hiltner1949science}
was studied in detail in 175 stars with a \mbox{$D=1.02$~m}
telescope \cite{hall1949aj}. This study was then followed by a
high-precision photometric survey of 841 stars carried out over
1.5 years~\cite{hiltner1951apj}.

 \item There is also one argument that concerts just spectroscopy on small telescopes. The fraction of high- and
medium-resolution spectroscopic studies will increase because of
the ever increasing light pollution in observatories founded
 in the XIX--XX~centuries. Moving (mostly photometric) small telescopes to sites located far from large cities and operating them
there is impractical. Whereas in the \mbox {1960--1970s} the
European Southern Observatory (ESO) was to a considerable extent
equipped with such telescopes, by the end of the century these
instruments, which became viewed as intermediate and small, were
decomissioned by the international organization and are now again
operated at the expense of their national owners. Such telescopes
can prove to be of use for spectroscopic studies.
\end{list}

 \item \underline{ Psychological:}
\begin{list}{$\bullet$}{
\setlength\leftmargin{4mm} \setlength\topsep{2mm}
\setlength\parsep{0mm} \setlength\itemsep{2mm} }
 \item There are productive experts in any field of science and they view their creative research as largely
colored by individualism. However, technically sophisticated
experiments on large telescopes are collective work. The
availability of small telescopes to a certain degree solves the
psychological problems of the researchers who by no means view the
individual nature of the process of astronomical observations as a
secondary factor.

 \item For many observing astronomers the priority of performing their own (sometimes the first) observation of a given
object (or phenomenon) is an important factor compared to a study
based on archival data.

 \item Small telescopes are better suited for educational purposes. Here the major educational factor is
the ability to influence the observing process (there are a lot of
remotely controlled training telescopes). Small telescope do not
develop separately from large  telescopes because the progress in
the use of small telescopes is entirely due to the progress in
signal detection and the emergence of fundamentally novel optical
methods. New technologies first appeared on large telescopes,
although there are quite a few exceptions. What was considered
large telescopes  just fifty years ago are now viewed as
moderate-diameter instruments. We will therefore consider the
problem of the equipment of small telescopes also in from
historical perspective. There is a huge body of literature on the
subject. Our aim was not to provide a full review, we limit
ourselves to outlining the examples, which we considered to be
either typical or of additional interest for us. We group the
kinds of equipment and methods by the type of detectors employed,
and within the same detector type, by the specificities of design
solutions.
\end{list}
\end{list}

\section{Spectrographs  with photographic registration}
Most of the technical solutions in the design of stellar
spectrographs date back to the epoch of photoghraphic registration
of spectra. Some spectrographs were later reequipped with
detectors of a new type. Moreover, some of the well-proven
classical schemes can be revisited with increasing CCD format.

\subsection{Prismatic spectrographs}
The advantages of prismatic schemes include the concentration of
the spectrum within a single strip, and its drawbacks, the
temperature dependence of the acquired spectrum (which is stronger
than in the case of diffraction spectrographs). Restricting
factors also include the size of the glass block with high
homogeneity and uniformity reqruirements to be met within its
volume.

\subsubsection{Slitless prismatic spectrographs}
The role of the focal ratio of the spectrograph lens was
demonstrated already during the establishment  the Harvard
classification: observations with the Bache telescope ($D=0.2$~m,
\mbox{$1:5.6$}) \cite{pickering1891annals} equipped with an
objective prism (or a set of prisms) made it possible to acquire
stellar spectra that were beyond the reach of slit prismatic
spectrographs used with much larger telescopes with smaller focal
ratios.

To classify spectra by their fragments near the Balmer limit, hot
stars were observed on Jungfraujoch observatory (Switzerland)
located at an altitude of  3457~m \cite{arnulf1936journal}. A
camera ($D=0.4$~m, $1:1.5$) with a quartz objective prism was
employed and the spectra were broadened via artificially produced
astigmatism (the lens was inclined by $\arcdeg{8}$ with respect to the
rear plane of Cornu prism).

The invention of Schmidt camera opened up, in particular, the
possibility to photographically record spectra of extended
objects. A two-prism nebular spectrograph with a Schmidt camera
was attached to the lower part of the Yerkes refractor, whereas
the entrance slit was placed in the upper part so that the
telescope ($D=1.03$~m) served both as a guide and support
structure for the nebular spectrograph whose field of view was
determined by the diameter of the Schmidt camera and the length of
the refractor tube \cite{struve1937apj}. This experiment was
further developed in the design of the nebular spectrograph of
McDonald Observatory~\cite{struve1938apj}, where terrain
configuration made it possible to position the entrance  ``slit''
far from the prismatic dispersing unit at a distance much greater
than the length of the Yerkes refractor length. the idea of taking
advantage of the terrain configuration was also implemented in the
scheme of the nebular spectrograph of Crimean Astrophysical
Observatory designed by D.~D.~Maksutov and B.~K.~Ioannisiani and
installed at a mountainside near Simeiz  (see
Pikelner~\cite{pikelner1954ika}). The main parts of the spectrograph were
a meniscus camera \mbox{($D=0.15$~m, $1:1$)} and two flint glass
prisms. Note that in the above design solutions spectra of stars
located in the field of view of the nebular spectrograph were
below the photographic registration threshold.

A slitless quartz spectrograph was used for spectrophotometry of
hot stars near the Balmer discontinuity
\cite{mirzoyan1955photometric}. An afocal $D=0.25$~m telescope
was used to do without expensive objective prism. The quartz
spectrograph provided a reciprocal linear dispersion of
\mbox{$P=150$~\AA/mm} near the H$\gamma$ line, and the spectrum of
a  $m_V=7^{\rm m}$ star could be acquired in a one-hour long
exposure. The study investigated  \mbox {OB-stars} in the Cep\,II
association.

The main shortcoming of prismatic slitless spectra is the lack of a comparison
spectrum. An attempt  to address the problem was undertaken long ago by
Pickering~\cite{pickering1896an}, who proposed reversal of the objective prism.
However, a comparison of two mutually reversed spectra does not yield pure
Doppler shift because: (a)~in the case of prism reversal the centers of the
direct and reversed exposures have different declinations (the centers should
coincide to within better than~$\arcsec{0.05}$) and (b) prism distortion
produces an extra shift of lines, which in the case of a two-degree field of
view results in errors as large as several thousand km\,s$^{-1}$. The former
problem restricts the application of the method to measuring relative radial
velocities, and the latter was overcome by Fehrenbach~\cite{fehrenbach1947cnd}.
Fehrenbach prism is a plane-parallel plate consisting of two prisms made of
different kinds of glass. The prisms have different dispersions but identical
refractive indices for a certain wavelength. Between two consecutive exposures
taken with the same plate the prism is reversed by~$\arcdeg{180}$ about the
telescope axis and slightly shifted in right ascension. The most productive
instruments for such system proved to be the GPO (Grand Prism Objectif, in
French) astrograph (with $D=0.4$~m, $1:10$, and the image scale of
$\arcsec{51.5}$~mm$^{-1}$), which provided the reciprocal linear dispersion of
$P=110$~\AA/mm \cite{gieseking1979st, gieseking1979tm}, and the Schmidt
telescope of  the Observatoire de Haute-Provence (OHP) ($D=0.62$~m, $F=2.23$~m,
$P=200$~\AA/mm at \mbox{$\lambda=4220$~\AA).} Observations on the GPO astrograph
with $2\times30$~min exposures yielded radial velocities $V_r$ with errors
between 4 and 9~km\,s$^{-1}$ for stars brighter than $\m{9.7}$. A comparison
with slit spectroscopy on a  $D=0.9$~m reflector showed the GPO to be five times
more efficient in per star terms if the differences in the aperture are taken
into account. There were a total of 35 program stars inside the field of view of
the telescope and therefore the total gain of GPO was a factor of 150--200
compared to slit spectrography on a telescope of the same aperture provided the
same, albeit rather low requirements in terms of the precision of  $V_r$. The
average error of the  Fehrenbach and Burnage~\cite{fehrenbach1981aaa} catalog
based on OHP observations is  4.2~km\,s$^{-1}$.

\subsubsection{Prismatic slit spectrographs}
Because of mechanical and temperature deformations prismatic
spectrographs with high reciprocal linear dispersion $P$ have not
become popular instruments for tasks involving Doppler
measurements. Low  $P$ ranging from 130 to 40~\AA/mm were used for
spectral classification. The foundations of two-dimensional MKK
spectral classification \cite{morgan1943atlas} were established
based on observations made with Yerkes refractor ($D=1.03$~m)
equipped with a single-prism spectrograph (its slit-width factor
is equal to 7) designed by G.~Van~Biesbroeck. The spectrograph
recorded the wavelength interval \mbox{3920--4900~\AA,} with
\mbox{$P=120$~\AA/mm} at H$\gamma$.

By mid-XX century prismatic slit spectrographs on telescopes with
aperture diameters \mbox{$D<1$~m} were used only in three cases.
On Lick Observatory a three-prism spectrograph
\cite{campbell1898mills} was used on the refractor ($D=0.91$~m)
and a two-prism spectrograph \cite{mayall1936pasp}, on the
Crossley reflector  ($D=0.91$~m). First, a single-prism
spectrograph and then, since 1927, a Curtis spectrograph
(collimator focal distance \mbox{$1:18$,} $F_{\rm {coll}}=68$~cm,
two 60-degree flint-glass prisms, interchangeable cameras with
focal distances \mbox{$F_{\rm {cam}}=7.5, 15, 30$}, and $60$~cm
and a set of reciprocal linear dispersions $P=140, 76, 38$, and
$19$~\AA/mm, respectively, were used on the reflector ($D=0.94$~m)
of Ann Arbor Observatory (in the outskirts of Detroit). The
spectrograph had been operated for more than 30 years. These
spectrographs were also used mostly for two-dimensional spectral
classification~\cite{morgan1943atlas, titus1940apj}.

Simeiz reflector ($D=1.0$~m) was equipped with a single-prism
spectrograph with an $F_{\rm {cam}}=55$~cm camera (its wavelength
range spanned from 3600~\AA{} to H$\alpha$, \mbox{$P=36$}~\AA/mm
at H$\gamma$) mounted in accordance with the scheme ``folded
Cassegrain'' scheme with a plane diagonal mirror ($1:18.6$)
\cite{albitzky1932radvel}. The prismatic spectrograph based on
V.~A.~Albitzky's design and used with the
 \mbox{$D=1.22$}~m reflector \cite{kopylov1954ika} had a collimator focal distance of $F_{\rm
 {coll}}=99.5$~cm
with a collimated-beam diameter of $d=5$~cm, a~$\arcdeg{66.6}$
flint-glass prism, and three interchangeable lenses ($1:4$, $1:8$,
$1:12$) with the focal distances of \mbox{$F_{\rm {cam}}=23$},
$48$, and $72$~cm, respectively. These lenses provided the
dispersion of  \mbox{$P=72, 36$}, and $23$~\AA/mm, respectively
(at the H$\gamma$ line). The two-prism quartz spectrograph, which
was also used on the $D=1.22$~m telescope of Crimean Astrophysical
Observatory, had the collimator focal distance or $F_{\rm
{coll}}=80$~cm with the collimated-beam diameter of $d=4$~cm, and
the camera focal distance of \mbox{$F_{\rm cam}=20$~cm ($1:4$).}
The optics of the camera allowed the spectrograph to operate in
the  \mbox{3400--4300~\AA} wavelength interval with a dispersion
of \mbox{$P=[65; 162]$~\AA/mm}, respectively.

A single- and four-prism spectrographs were used in the Newton
focus of the telescope ($D=1.2$~m) of Saint-Michel Observatory. A
total of more than 4500 spectra including 148 spectra of Be stars
were acquired over 30 years since 1944 on the single-prism
spectrograph ($F_{\rm {coll}}=400$~mm, $d_{\rm {coll}}=70$~mm,
$F_{\rm {cam}}=215$~mm)~\cite{hubert1979atlas}.

The spectral classification of hots stars by the fragment of the
spectrum near the Balmer jump was based not only on spectra
acquired with the prismatic camera~\cite{arnulf1936journal}, but
also on spectra acquired with Chalonge's two-prism slit quartz
spectrograph~\cite{baillet1952recherches}, $F_{\rm
{coll}}=330$~mm, and two Cornu prisms, $F_{\rm {cam}}=118$~mm.
Spectroscopic observations on $D=0.25$ and $0.8$~m telescopes
recorded the wavelength interval from 3100~\AA{} to H$\alpha$.
Nonstandard motion of the cassette was used to broaden spectra to
0.35~mm at 3100~\AA{} and 1.5~mm near H$\alpha$.

\subsection{Diffraction spectrographs}
Diffraction spectrographs became competitive after the development
of ruled gratings with profiled grooves \cite{wood1935phrev}.
Further improvement of ruling machines~\cite{harrison1949prod,
gerasimov1957delit,gerasimov1957opyt} and manufacturing
technologies for diffraction gratings~\cite{kossova1958opyt}, as
well as the introduction of holographic techniques made it
possible to efficiently  use slit and slitless schemes for
diffraction systems. Profiling the grooves of diffraction gratings
increased substantially the angular dispersion of spectrographs
with the size of the dispersing element remaining the same.

\subsubsection{Slitless diffraction spectrographs}
Epstein~\cite{epstein1967pasp} proposed to combine a reflective Schmidt
corrector plate with a diffraction grating ($D=0.15$~m, $F=61$~cm,
\mbox{$P=150$}~\AA/mm). This idea was further developed in the
optical layouts of fast anastigmats \cite{lemaitre1976reflective}
and slit spectrographs \cite{lemaitre1981proc, lemaitre1983proc}
attached to telescopes of various diameters including instruments
with $D=1.0$ and $1.2$~m~\cite{fehrenbach1981observations}.

Hoag and Schroeder~\cite{hoag1970pasp} used a transparent diffraction grating in
the converging beam on the Kitt Peak Observatory reflector \mbox{($D=1.0$~m,}
$1:7.5$). In that case images of other spectral orders are located along a
circle whose radius is equal to the distance from the grating to zero-order
image. A 150~grooves/mm grating placed 5.1~sm from the focus provided a
dispersion of 1260~\AA/mm, with a working field of view having a $\arcmin{30}$
size. A half-hour exposure allowed registering the spectra of objects as faint
as $m_B=\m{16.8}$.

Linnik~\cite{linnik1963nta} proposed a solution to the problem of
simultaneous acquisition of the standard and stellar spectra with
a slitless spectrograph. A small part of the collimated beam
passes through the plate that forms Talbot's interference fringes
located in the focal plane of the lens next to the spectrum of the
star. The diffraction grating with a fixed diffraction angle has a
reflecting prism mounted on it with a small angle deviating a
small part of the collimated beam (without diffraction) to the
camera field of view, where this undispersed radiation is
deflected to the guide optics. Guiding is performed by rapidly
moving the first lens of the collimator across its primary axis.

\subsubsection{Slit diffraction spectrographs}
The main advantage of slitless spectroscopy~--- the capability to
simultaneously record several objects~--- goes along with a number
of shortcomings with the most important being the influence of sky
background and dependence of resolution of seeing and guiding
quality. Therefore the main efforts were channeled to improve slit
diffraction spectrographs.

The spectrograph of the first KPNO telescope ($D=0.91$~m) was used
for spectral classification~\cite{abt1963apjs}. A spectrograph
with a ``folded Schmidt'' camera ($1:2$)~\cite{meinel1963al} was
designed for the  $D=0.91$~m telescope of Steward Observatory. It
was equipped with a set of interchangeable gratings, which
provided a \mbox{$P=[22; 800]$~\AA/mm} set of reciprocal linear
dispersion values.

A combination of a diffraction grating with a Schmidt camera
changed the appearance of a slit spectrograph. The pursuance of
universal capabilities gave rise to the development a set of
photographic cameras often equipped with interchangeable
diffraction gratings incorporated into a single structure. This
resulted in the increase of the size and mass of equipment and
that is why such devices were used on  $D\ge1.5$~m telescopes. As
an example of such a solution we can mention the three-camera
Cassegrain spectrograph of the Mount  Wilson Observatory telescope
($D=1.5$~m) equipped with three interchangeable
gratings~\cite{wilson1956pasp}.

The focus of the  $D=1.2$~m Newton telescope of Saint Michel
Observatory had three separate adapters, which allowed
simultaneous observations to be performed using three techniques.
One of the techniques was represented by diffraction spectrograph
``E''~\footnote{\protect\url{http://www.obs-hp.fr/histoire/120/spectro_E.shtml}}
equipped with interchangeable  300 and 600~grooves/mm gratings and a
$f/2.4$ ($P=275$~\AA/mm) dioptric camera. The spectrograph was
also used with an image tube. Since 1959 the telescope has been
equipped with ``E${}^\prime$
spectrograph''~\footnote{\protect\url{http://www.obs-hp.fr/histoire/120/spectro_Eprime.shtml}}
manufactured by REOSC and meant for operating in the near IR
($P=230$~\AA/mm,  \mbox{I-N) emulsion.} The ``E${}^\prime$''
spectrograph was refitted with gratings and camera (``semi-solid
Schmidt camera'', $f/0.47$, \mbox{$P=290$~\AA/mm} and $f/2.5$,
$P=64$~\AA/mm).

The large prismatic ``glass'' spectrograph of the $D=1.22$~m
reflector of Crimean Astrophysical Observatory was in 1961
replaced by the first domestic-made ASP-11 diffraction
spectrograph. The spectrograph was equipped with two
interchangeable diffraction gratings, making it possible to
acquire spectra in two orders: in the 4800--6700~\AA{} wavelength
interval with a dispersion of 37~\AA/mm in the first order and in
the \mbox{3600--4800~\AA{}} wavelength interval with a dispersion
of  15~\AA/mm in the second order~\cite{rachkovskaya2013ika}. As
a result of such replacement the accuracy of radial-velocity
measurement degraded  \cite{chentzov2013ika}. When operating in
the near IR the spectrograph was equipped with a 300~grooves/mm
grating and an \mbox{FKT-1A} image tube
\cite{vitrichenko1975infrared}.

Mass instruments of the photographic epoch included  Boller \& Chivens~(B\&C)
and Karl Zeiss Jena spectrographs operated in the Cassegrain focus. The B\&C
spectrograph has a collimated beam diameter  $d=9$~cm, interchangeable
$102\times128$~mm$^2$ gratings, and a semi-solid Bowen-Schmidt-Cassegrain camera
\mbox{($F=14$~cm)} with  $6\times25$~mm$^2$ unvignetted field. UAGS spectrograph
has similar parameters of the external-focus ($F=15$~cm), two Schmidt cameras
with internal focus ($F=11$~cm and \mbox{$F=17$}~cm), but a smaller
collimated-beam diameter ($d=7.5$~cm) and more optical elements. These
instruments serves as a basis for transition of moderate-resolution spectroscopy
to photoelectric detectors. Spectroscopists in our country used UAGS universal
slit spectrograph commercially produced by Karl Zeiss Jena. In particular, the
spectroscopy technique involving the use of image tube was tested on the
\mbox{$D=0.6$~m} telescope of the Special Astrophysical Observatory of the
Russian Academy of Sciences and then used on the 6-m telescope as the main
method for the study of galaxies~\cite{afanasiev1981opyt}. UAGS set up for
photographic registration was used on the same telescope to monitor emission-
and absorption-line spectra of Mira variables~\cite{bychkov1978rapid,
morozova1978some}. Based on the results of observations on a $D=0.6$~m telescope
equipped with UAGS spectrograph ($P=28$~\AA/mm) Gulyaev et.~al~\cite{gulyaev1986determination}
developed a system of Balmer-line indices tied to stellar model atmospheres.

Richardson and Brealey~\cite{richardson1973jrasc} developed a small-sized photographic
spectrograph with an off-axis camera and collimator for the
Nasmyth focus of the \mbox{$D=0.9$}~m reflector. This instrument
was later equipped with a Reticon detector~\cite{edvin1989obs}.

\subsubsection{Diffraction spectrographs for coud\'e focus}
Before the introduction of echelle spectrographs operating at high
diffraction orders the primary way to increase the spectral
resolution was to increase the focal distance of the spectrograph
camera. Therefore the static coud\'e focus, which was introduced
as early as XIX century for visual observations (Paris
Observatory, 1882, the $D=0.27$~m refractor), was also used in the
XX century on small- and medium-diameter telescopes
(Table~\ref{tab_1_panchuk}). The parameters of the telescopes can
be found in our earlier review~\cite{panchuk2013high}.

\begin{table}
\caption{Some of the small- and medium-diameter telescopes used
for spectroscopy \newline and coud\'e spectroscopy} \label{tab_1_panchuk}
\medskip\centering
\begin{tabular}{c|c|c|l}
    \hline
Year & $D$, m & $D:F$ & \multicolumn{1}{c}{Observatory} \\
\hline
1922 & 0.91 & 1:36  & Steward            \\
1955 & 0.91 & 1:30  & Cambridge          \\
1963 & 0.91 & 1:37  & KPNO coud\'e feed  \\
1969 & 0.6  & 1:36  & Lick CAT           \\
1970 & 1.0  &       & Canopus Hill       \\
1970 & 0.61 & 1:28  & Fick, Iowa         \\
1971 & 0.4  & 1:33  & Canopus Hill       \\
1971 & 1.06 & 1:49  & Lowell             \\
1981 & 1.4  & 1:120 & ESO CAT            \\
1982 & 0.5  & 1:13  & Crimean AO         \\
1990 & 1.0  & 1:36  & SAO RAS${}^*$    \\
  \hline
\multicolumn{4}{l}{\parbox{7.5cm}{\strut\footnotesize *~-- The effective diameter of the telescope
was  $D<1.0$~m because of design errors in the coud\'e path optics.}}\\
\end{tabular}
\end{table}

\subsubsection{Diffraction-based cross-dispersion slit spectrographs}
The development of the technology of stepped-groove
gratings~\cite{harrison1949prod2, gerasimov1958izgot} made it
possible to concentrate the flux within a narrow interval of
diffraction angles. In the 1960s two-mirror schemes with coma
compensation across a sufficiently large field of view of the
camera were developed to operate in high diffraction
orders~\cite{kopylov1965ika, schroeder1967echelle}. The
parameters of the collimator should not  have differed to much
from those of the camera and therefore these solutions proved to
be optimal just for small telescopes, where small slit width
factor was quite acceptable. In the next 10 years the
spectrographs listed in Table~\ref{tab_2_panchuk} were put into
operation.

\begin{table*}
\caption{Cross-dispersion spectrographs in the Cassegrain focus}
\label{tab_2_panchuk}
\medskip\centering
\begin{tabular}{c|c|c|c|c|c|l}
        \hline
Year & $D$, m & $d$, cm & $\tan\theta_b$ & disp & $R$ & Observatory\\
    \hline
1971&  0.9 &  5.5& 2  &  ech/gr   & 16\,000  &  Pine Bluff Obs.~\cite{schroeder1971echelle} \\
1976&  0.91&  5  & 2  &  pr/ech   & 40\,000  &  Goddard SFC~\cite{mcclintock1979hires}      \\
1977&  0.61&  9  & 2  &  ech/gr   & 43\,000  &  Mt. John Obs.~\cite{hearnshaw1977casseg}    \\
1978&  0.9 &     & 2  &  pr/ech/pr& 40\,000  &  Royal Greenwich~\cite{mckeith1978cassegr}   \\
1978&  1.0 &  *  & 2  &  ech/gr   & 52\,000  &  Ritter Obs.~\cite{latham1977inastro}        \\
1980&  1.0 &  *  & 2  &  ech/gr   & 52\,000  &  Lowell Obs.~\cite{latham1977inastro}        \\
1980&  1.0 &     & 2  &  ech/gr   & 30\,000  &  Siding Spring Obs.                           \\
1981&  1.0 &  7.7& 2  &  ech/gr   & 54\,000  &  Vienna Obs.~\cite{weiss1981pasp}            \\
1982&  0.61&  *  & 2  &  ech/gr   &          &  Las Campanas~\cite{latham1977inastro}       \\
1982&  0.61&  5  & 3.2&  filt/ech & 150\,000 &  Whipple Obs.~\cite{hunten1991cassegr}       \\
1986&  1.22&     & 2  &  ech/gr   & 50\,000  &  Rangapur Obs.                                \\
    \hline
\multicolumn{7}{l}{\footnotesize {Designations: disp~-- sequence
of dispersing elements  along the ray path, ech~-- echelle,}}\\
\multicolumn{7}{l}{\footnotesize {gr~-- grating, pr~-- prism, filt~-- filter; $R$~-- spectral
resolution, $D$~-- diameter of the mirror,}}\\
\multicolumn{7}{l}{\footnotesize {$d$~-- diameter of the collimated beam. *~-- copies of  Harvard
College Observatory spectrograph,}}\\
\multicolumn{7}{l}{\footnotesize { used on the $D=1.52$~m telescope}}\\
\end{tabular}
\end{table*}

For over 20 years, until the beginning of the  ``fiber-optics
era'', echelle spectrographs were used for stellar spectroscopy on
small telescopes. This was because of the gain of
spectral resolution, which is proportional to the tangent of the blaze angle,
making it possible to achieve high and moderate spectral
resolution using rather compact suspended systems. The development
of these systems was hampered by problems with digitizing of
photographic echelle spectra with microdensitometers (the use of a
prism as a cross-dispersion element resulted in the curvature of
spectral orders) and bright limiting magnitude (spectra of stars
brighter than sixth magnitude were recorded).  Some of the echelle
spectrographs were later used with image tubes.

\section{Spectrometers  with single-channel photoelectric registration} 
Single and two-channel spectrometers have been the main detectors
of small telescopes for about thirty years.

\subsection{Slitless scanner with an objective prism}
For spectrophotometric observations in the near IR a catadioptric
telescope based on a design by P.~P.~Argunov ($D=0.43$~m, $1:10$)
\cite{argunov1967invest} was used with a four-degree objective
prism and  RCA-7102 photomultiplier. Dispersion was directed along
declination and the spectra were scanned (4000--10\,000~\AA{} in
10 minutes) with a reversible motor
\cite{komarov1968azh}.

\subsection{Prismatic scanner on an afocal telescope}
ASI-5  afocal Mersenne telescope ($D=0.25$~m) was used for
photoelectric scanning of stellar spectra \cite{melnikov1956azh}.
A parallel beam formed after the reflection from the parabolic
secondary convex mirror of the telescope hit a 30-degree Littrow
prism, and the dispersed beams were then caught by a concave
mirror, which provided a dispersion of  200~\AA/mm at H$\gamma$.
The spectrum was scanned by turning the prism, and the
14~\AA{}-wide measured spectral portion passed through the slit
with a photoelectric photometer. The signal detection and scanner
control system included 17 radio lamps. The device was highly
efficient in the UV and allowed acquiring spectra of stars down to
the seventh magnitude~\cite{melnikov1959absolute}.

\subsection{Scanning monochromators}
To change the spectral resolution on a photographic spectrograph one has to
change the focal distance of the camera. Jacquinot and
Dufor~\cite{jacquinot1948krcnrs} showed that spectral resolution achieved on
monochromators depends on the slit width, i.e., monochromators are more flexible
spectroscopic devices compared to photographic spectrographs. The main
shortcoming of scanning monochromators is that seeing and transparency
fluctuations at the spectrometer entrance show up in the details of the spectra.
Hiltner and Code~\cite{hiltner1950josa} proposed a method for compensating
fluctuations by comparison with the signal from the reference channel, which
records the fraction of the flux that passed through the entrance slit.

\subsubsection{Prismatic slitless monochromators}
A system for compensating fluctuations of the illumination of the
entrance slit was used in the two-prism monochromator mounted on
the  $D=1.2$~m reflector \cite{geake1956mnras}. To achieve
maximum resolution (0.5~\AA) the slit width had to be reduced to
25~$\mu$m, whereas the size of the stellar image and the amplitude
of image tremor were much greater. At the time, a complex system
for fluctuation compensation was used: the photoelectric
multiplier of the second channel recorded undispersed light
reflected from the first face of the prism.

\subsubsection{Scanners with a flat diffraction grating}
Boyce et.al.~\cite{boyce1973pasp} developed a scanner for coud\'e focus of the
$D=1.06$~m telescope of Lowell Observatory. The collimator used
had a focal distance of  \mbox{$F_{\rm{coll}}=610$~cm,}
collimated-beam diameter $d=12.7$~cm, 1200~grooves/mm grating,
camera focal distance  $F_{\rm{cam}}=305$~cm, reciprocal linear
dispersion 2.67~\AA/mm at 5000~\AA, and a scanning step of
0.078~\AA{} or more. The device was used to study axial rotation
velocities of bright stars.

The scanner of the $D=0.61$~m telescope of Bochum University is a
good example of a device that was used efficiently to measure
spectral energy distributions in stars of various types. In 1968
the telescope was moved to the European Southern Observatory where
it was used to establish spectrophotometric standards for the
Southern sky. The single-channel instrument was made according to
the Czerny--Turner scheme, and scanning of the spectrum was
achieved by moving the grating \cite{dachs1976messenger}.

\subsubsection{Scanners with a concave grating}
Liller~\cite{liller1963ao} showed that the transmission of a monocromator with a
concave grating increases by a factor of three compared to a monochromator with
a flat grating. The most popular among various monochromator schemes with
concave gratings became the one proposed by Namioka~\cite{namioka1958josa},
where the angle between the lines connecting the grating center and the slits,
as well as the distance between the grating and the slits remain unchanged. The
most productive spectrophotometer proved to be the device based on this scheme
and made for the $D=0.5$~m reflector~\cite{kalinenkov1967afi}; with this device
spectrophotometry of stars down to the seventh magnitude could be performed
\cite{haritonov1972sao}. For observations on a $D=0.37$~m telescope Beavers and
Eitter~\cite{beavers1986instr} used a 1200~grooves/mm concave ($1:4$)
holographic grating providing a reciprocal dispersion of 40~\AA/mm.

\subsection{Narrow-band spectrophotometers}
The most obvious method for line spectrophotometry is to switch a system of
slits centered on the measured line (group of lines) and onto the neighboring
fragments of the continuum. Gustaffson and Nissen~\cite{gustafsson1972metal}
used an echelle spectrograph providing high dispersion ($P=2$~\AA/mm). In their
program of the study of helium content \cite{nissen1974hetohy} used a
14~\AA{}-wide slit centered onto the 4026~\AA\ line and two  6~\AA{}-wide slits
serving to measure the continuum flux on both sides of the helium line. With up
to 100000 counts per half-hour 10th magnitude stars could be observed on the 1-m
telescope of European Southern Observatory~\cite{nissen1977messenger}. In the
Northern Hemisphere the spectrophotometer was operated on the  OHP telescope
($D=1.93$~m).

The photomultiplier sizes made it impossible to use multichannel
systems, which had proved to be a good choice for large
(Oke~\cite{oke1969pasp}, Rodgers et.~al.~\cite{rodgers1973pasp}, etc.) and
moderate-diameter telescopes. The introduction of miniature
photoelectric multipliers and optical-fiber technology allowed
solving this problem~\cite{barwig1986instrumentation}. The
object, comparison star, and sky background are projected
simultaneously onto the three 400-micron entrance fibers of the
multichannel spectrophotometer.  The offset guide allows objects
as faint as $16^{\rm m}$ to be observed with a 1-m telescope.
Optical fibers feed three identical prismatic spectrographs. The
spectrum produced by each of the three spectrographs is projected
onto the faces of 15~optical fibers whose output ends are
connected to Peltier-cooled miniature photoelectric multipliers.

\subsection{Correlation spectrometers}
Griffin~\cite{griffin1967apj}, whose used the coud\'e focus of the
\mbox{$D=0.91$~m} telescope, was the first to practically demonstrate the
efficiency of cross-correlation technique for measuring radial velocities.
Griffin's technique of radial-velocity measurements on small telescopes gave a
three orders of magnitude gain compared to the photographic
method~\cite{griffin1969photoelectric}. A description of the \mbox {$D=0.61$}~m
telescope of Fick Observatory dedicated for radial-velocity measurements can be
found in Beavers and Eitter~\cite{beavers1977pasp}. A mirror with a diameter of
$d=41$~cm and a focal distance of \mbox{$F=305$}~cm, which serves as a
collimator and a camera, is mounted in the cylindrical camera of the coud\'e
focus. The 1200~grooves/mm diffraction grating (with maximum concentration at
5000~\AA\ and the size of the  \mbox {$135\times110$}~mm$^2$ line area has its
lines oriented parallel to the North--South direction. The reciprocal linear
dispersion is 2.62~\AA/mm, and the slit mask covers a wavelength 400~\AA{}
wavelength interval centered on 4600~\AA. The 0.09~mm entrance slit corresponds
to an image size of $\arcsec{1}$ and resolution of 15~km\,s$^{-1}$ on the mask.
The cylinder where the spectrometer is mounted must have its air pumped out.
Wavelength calibration is based on the lines of the spectrum of a helium-neon
laser.

Practically at the same time the first cross-correlation
spectrometer with echelle~--- CORAVEL~--- was developed
\cite{baranne1979vistas}. The first such device operated since
1977 at the $D=1.0$~m Swiss telescope of Observatoire de Haute
Provence (OHP)~\cite{baranne1977sur}. It allowed determining the
radial velocity of a \mbox{$m_B=11^{\rm m}$} K-type star with a
standard deviation of 0.7~km\,s$^{-1}$ in 10 independent 0.5-min
exposure measurements. The second CORAVEL device was since 1981
used at the $D=1.54$~m Danish telescope of European Southern
Observatory. Because of its compact size of the device
\mbox{($F_{\rm{cam}}=57$~cm)} the device was quite rigid and
convenient to use in the Cassegrain focus. The 0.5~km\,s$^{-1}$
accuracy of radial-velocity measurements was achieved for
$m_B<15^{\rm m}$ stars~\cite{imbert1981first}.

An efficient photoelectric radial-velocity meter based on a
coud\'e spectrograph~\cite{richardson1968} was developed for the
$D=1.2$~m telescope of DAO \cite{fletcher1982pasp}. Its
parameters are: a $F_{\rm{coll}}=757$~cm ($1:30$) collimator,
mosaic 831~grooves/mm diffraction grating,  $F_{\rm {cam}}=244$~cm
camera, reciprocal linear dispersion \mbox{$P=2.4$~\AA/mm,}
0.08~mm slit projection width, a mask moved by a step motor, and a
($1:1$) plastic Fabry lens mounted in front of the photoelectric
multiplier. With 10 counts per second made for a $m_B=16^{\rm m}$
star the final accuracy was better than 1~km\,s$^{-1}$.

It is interesting that the cross-correlation technique was also
used to study dust particles in the F
corona~\cite{beavers1980aj}. A commercially produced Ebert
monochromator ($1:3.5$) was equipped with a slit mask reproducing
the locations of the specially chosen lines in the solar spectrum
in the 4005--4352~\AA\ wavelength interval. A $D=0.15$~m ($1:5$)
parabolic mirror combined with a flat coelostat mirror was used as
feeding optics.

The small spectrograph mentioned above and designed for
photographic observations \cite{richardson1973jrasc} was
reequipped into the cross-correlation spectrometer of the
$D=0.4$~m DAO telescope. A special prism was mounted behind the
slit and a mask (300 lines in the 415--485~nm wavelength interval)
based on the spectrum of a K-type giant was installed in the focal
plane of the camera. The accuracy of radial-velocity measurements
(2~km\,s$^{-1}$) was limited by the insufficient rigidity of the
small device.

Domestic correlation radial-velocity
meter~\cite{tokovinin31stellar}, which has been used with
$D=0.6$, $0.7$, $1.0$, and $1.25$~m telescopes since 1984 is
currently one of the most long-lived devices based on this method.

\subsection{Spectrum coding systems}
The development of spectrum coding devices was sort of a crowning
moment in the history of single-channel photoelectric systems. The
method of spectrum coding consists in the registration by a
single-channel detector of a sequence of signals produced as a
result of the passage of light through a successfully changed slit
masks. The sequence of slits and nontransparent parts of the mask,
as well as the sequence of the replacement of such masks, can be
made according to a certain pattern to make the digital processing
of the set of signals a sufficiently simple process. For example,
the replaced slits of the mask can be arranged in accordance with
the algorithm of the construction of strings in the Hadamard
matrix \cite{sloane1969ao}. The interest in experiments with
Hadamard spectrometers is easy to explain: they offer the same
multiplex gain as a Fourier spectrometer (i.e., the $S/N$ is
proportional to the square root of the number of discrete elements
of the spectrum), whereas the technology is much simpler: a
diffraction spectrograph with mechanically changed masks is used.
In the case of extended objects a combination of coding masks at
the entrance to and exit from the spectrometer can be used
\cite{harwit1971ao}. An Hadamard spectrometer allowing
simultaneous acquisition of 15 \mbox {IR spectra} each consisting
of 255 elements \cite{minghingtai1975ao} was used to study the
latitudinal distribution of methane across the disk of Jupiter.

\section{Spectrographs  with multichannel photoelectric  detection}   
The introduction of multichannel detectors into astronomical
spectroscopy made it possible: (a)~to extend the observed
wavelength range; (b)~to increase the spectral resolution, and
(c)~to gradually increase the photometric accuracy. Practically
all types of multichannel detection came to be used on telescopes
of diameters considered.

\subsection{Spectrographs with image tubes}
Most of the spectrographs with photographic registration whose
design allowed the use of lens cameras or Schmidt--Cassegrain
cameras served as a basis for testing observational techniques
involving the use of image tubes with subsequent photographic
registration. For example, the UAGS spectrograph equipped with an
image tube~\cite{afanasiev1981opyt}, which was developed for the
6-m telescope of the Special Astrophysical Observatory of the
Russian Academy of Sciences, was first tested on the
\mbox{$D=0.6$~m telescope.} The use of an image tube with optical
transfer to a photographic plate yielded no strong gain in terms
of the  ``limiting magnitude for the given $S/N$'', and that is
why at first the main reason for the use of an image tube was to
explore the near \mbox{IR part} of the spectrum, which was
inaccesible for direct photographic observations. For example,
Wood~\cite{wood1968pasp} used the Cassegrain spectrograph of the
$D=0.81$~m reflector to acquire Zeeman spectra of bright stars in
the IR. Autocorrelation technique was adapted to search for lines
that are asymmetric because of the Zeeman or other effects.
Esipov~\cite{esipov1963novtech} designed an efficient instrument of this
kind. The spectrograph kit included lens cameras corrected for
operating in the IR and providing the dispersions of  90 and
200~\AA/mm with a 600~grooves/mm grating in the first order. The device
was tested on the \mbox{$D=0.33$~m reflector.} The very first
Cassegrain echelle spectrograph \cite{schroeder1971echelle} was
used on the $D=0.91$~m telescope including observations with an
image tube.

\subsection{Spectrographs with electronographic cameras}
The first experiments with an electronographic
camera~\cite{lallemand1956astronomie} were carried out on ``E''
spectrograph in the Newton focus of the  $D=1.2$~m telescope of
Saint-Michel Observatory. Because electronographic cameras are
difficult to operate in suspended arrangements (Newton and
Cassegrain foci), the most efficient camera for spectroscopy on
medium-diameter telescopes proved to be the one mounted in the
coud\'e focus of the  $D=1.52$~m telescope of European Southern
Observatory. ``ECHEL.E.C.'' (echelle spectrograph with an electron
camera)~--- the first universal white-pupil spectrograph~--- was
mounted there in the early 1970s~\cite{baranne1972auxinstr}.
It had a Lallemand--Duchesne electron camera with electrostatic
focusing and 30-mm S-11-type photocathode as a detector. When
operating in the ``echelle'' mode it provides a dispersion of
4.5~\AA/mm in the blue part of the spectrum. A spectrum of a
$m_B=10^{\rm m}$ star broadened to 0.2~mm was acquired in a
2.5-hour long exposure. When operated in the single-order mode
with a dispersion of  74~\AA/mm, the device achieved a limiting
magnitude of $m_B=14^{\rm m}$ in one-hour
exposure~\cite{breysacher1976messenger}. Electronographic
spectroscopy technique proved to be quite complex and that is why
the spectrograph also used photographic registration including
registration via an image tube. The white-pupil scheme was also
used in the design of a nebular spectrograph with an image tube
\cite{baranne1974preliminary}.

\subsection{Spectrographs with dissectors}
In the early 1970s an image dissector scanner (IDS) was used on the $D=1.0$~m
telescope of Lick Observatory. It consisted of a three-stage image tube (with a
40-mm diameter  S-20 photocathode), with a dissector scanning the spectrum
during the phosphorus afterglow placed behind the third stage. Thus the
phosphorescent screen was used as an intermediate memory device. The
spectrograph had two apertures with the second one used for registration of the
sky background. Aperture switching was used to more accurately take the sky
background into account. The peak quantum efficiency of a dissector with 350
resolution elements was 20\%. A shortcoming of dissector is the long afterglow
of luminophor in the case of overillumination. In observations of the peculiar
object SS\,433 an IDS was used that mounted on telescopes of various diameters
including the  $D=0.6$~m reflector of Lick Observatory~\cite{margon1979aj}.
McNall et.~al.~\cite{mcnall1972pasp} equipped echelle spectrograph
\cite{schroeder1971echelle} of the $D=0.91$~m with an image tube with a
dissector. The device registered 256 resolution elements in one of the orders of
the echelle spectrum. However, the dissector was not a photon counter.

\subsection{Spectrographs with diode arrays}
The spectrograph of the $D=0.9$~m  telescope of Kitt Peak
Observatory, which was operated since early 1980s, is a good
example of an intensified Reticon scanner (IRS). The dissector
after the image tube was replaced by two 820-element Reticon
arrays. Readout was performed every 10 seconds and each image
contained noise making up for four events per element. The
accuracy of radial-velocity measurements on such a device was
\mbox{3--5}~km\,s$^{-1}$.

A spectrograph for spectral classification in the near IR was
designed for the $D=0.91$~m, $1:10$ telescope of  MIRA
observatory ($h=1520$~m above sea
level)~\cite{torresdodgen1993atlas}. This Cassegrain-focus
spectrograph ($F_{\rm {coll}}=380$~mm, 600~grooves/mm grating,
$\lambda_{\rm{max}}=8400$~\AA, replaceable cameras with $F_{\rm
{cam}}$ ranging from 55 to 600~mm) was used with a Reticon array
whose high readout noise (no more than $1000\,\mathrm{e}^{-}$) is
not critical for spectrophotometry mode
\mbox{($2\times10^7\mathrm{e}^{-}$} per diode).

\subsection{Spectrographs with photon counters}
Schectman and Hiltner~\cite{schectman1976pasp} described a multichannel
spectrograph with a photon counter. The system includes two three-cascade image
tubes connected by a fiber disk with optical transfer to the diode array. With a
$D=1.3$~m telescope one count per second per Angstrom was obtained for a
$m_V=13^{\rm m}$ object. The ``shectography'' technique evolved into the
2D-Frutti system~\cite{schectman1984spie}. The device was developed for the
$D=4.0$~m CTIO telescope and then the same design was reproduced for a
$D=1.0$~m, $1:10$ telescope with a Boller \& Chivens spectrograph. It was a
two-cascade image tube with a  40-mm S-20 cathode connected via optical transfer
line to an electrostatically focused  40-mm image tube, which, in turn, was
connected with a microchannel-plate image amplifier via a fiber disk. The image
is then transferred to the CCD via a scale fiber transformer (phocon). The
output $35\times28$~mm$^2$ output  field of the spectrograph is then produced on
the $11.3\times8.8$~mm$^2$ CCD. The total CCD readout time is  7~ms, however,
one can read just a part of the CCD field. The total slit length, which has the
size of $\arcmin{6.8}$, occupies only 32 rows on the CCD. Repeated event counts
are struggled against at the stage of comparison with the preceding frame. Event
coordinates are determined to within $1/8$ of the pixel size. One thus obtains a
$3040\times256$~pixel frame at the output. The device is characterized by
nonlinearity at large fluxes (more than 9 counts per resolution element per
second). On a  $D=1.0$~m telescope this corresponds to registering a
$m_V=12^{\rm m}$ object with a resolution of 4~\AA. This system was more
expensive than CCDs manufactured in the early 1980s. The photon counter
developed by Mochnacki et.~al.~\cite{mochnacki1985ddo} for the $D=1.9$~m
telescope of DDO was tested on a \mbox{$D=1.0$~m telescope.} Observations with a
reciprocal linear dispersion of 16~\AA/mm provided a radial-velocity measurement
accuracy of 1~km\,s$^{-1}$ for stars down to a limiting magnitude of
$m_V=15^{\rm m}$. Thus the most complex systems of the registration of spectra
successfully operated on meter-class telescopes.

\subsection{Spectrographs with linear and 2D CCDs}
Reticon-type arrays had noise that was one order of magnitude
fainter than the noise of CCDs of the mid-\mbox{1980ies} and that
is why successful use of linear arrays was associated with the
possibility of acquiring long exposures. Efficient use of a
Reticon in the  coud\'e focus of the 1.2-m telescope of the
University of West Ontario (UWO) made it possible to develop
techniques for identifying the broadening components (rotation,
macro- and microturbulence) of absorption-line
profiles~\cite{gray1976theobs, gray1978high}. A 316~grooves/mm
R2-echelle was used with one of the orders \mbox {($m = [6; 15]$)}
separated via interference filters (with 85\% transmission). The
``folded Schmidt-type'' camera \mbox{($F = 559$~mm)} provided a
dispersion of  0.038~\AA/diode (6250~\AA, ninth order). With these
parameters a one-hour exposure yielded \mbox {$S/N = 100$} for a
sixth-magnitude star. Microscanning by shifting the detector unit
position by half the width of the 15-micron diode before repeating
the exposure was used to double the number of points on the
spectral-line profile. Doubling the number of data points on the
line profile formally facilitated the separation of broadening
components in the Fourier domain. In such studies one has to\ take
into account line broadening due to spectrograph flexibility. The
coud\'e focus spectrograph of the UWO telescope \mbox{($D=1.2$~m)}
was known for its high stability (dome temperature variations did
not exceed one degree per week). In the case of long exposures
lines can also broaden because of the Earth's diurnal rotation.
For example, when a zero-declination star is observed from the
equator diurnal rotation of the Earth broadens a line by
0.35~km\,s$^{-1}$ near meridian, and the effect is stronger at
other hour angles, whereas line broadening parameters have to be
known to within 0.1~km\,s$^{-1}$ in some cases
\cite{gray1986instrum}.

Recall that the CES coud\'e echelle spectrograph of European
Southern Observatory \cite{enard1979messenger} was mostly used
with the auxiliary $D=1.4$~m telescope
\cite{andersen1977messenger, andersen1979messenger} of the
coud\'e focus of the large $D=3.6$~m telescope. This spectrograph
was used to perform many programs requiring high spectral
resolution combined with low level of scattered light (as a result
of preslit filtering and the use of the double-path system). The
popularity of the device was also due to the fact that it allowed
observations to be remote-controlled from Europe.

Furenlid~\cite{furenlid1984pasp} proposed a scheme of the telescope-spectrograph
 with only one optical element~--- a concave diffraction
grating. Estimates showed that such a telescope with an equivalent
diameter of $D=0.18$~m, focal distance $F=3.3$~m, line density
210~grooves/mm, and a dispersion of $P=14.3$~\AA/mm provides spectra
of a $\m{10.6}$-magnitude star to be acquired  on a CCD with a
signal-to-noise ratio  $S/N=100$. The above authors pointed out,
for comparison, that the limiting magnitude of the coud\'e-focus
spectrograph with the same dispersion and $S/N$ ratio operated on
the \mbox{$D=0.97$~m }  KPNO telescope has 1$^m$ brighter limiting
magnitude. It goes without saying that a single-element
spectrograph telescope retains all the shortcomings of slitless
spectroscopy.

Further improvement of the CCD charge coupled device (CCD)
technology substantially facilitated designing suspended
spectrographs by reducing the lower limit for the diameters of
spectroscopic telescopes. McDavid~\cite{mcdavid1986instr} developed a
miniature ($0.3\times0.18\times0.13$~m$^3$) Cassegrain
spectrometer for a  $D=0.4$~ ($1:12$) telescope. Its detector
consists of a 128-element CCD array (each element has the
size of $13\times13$~$\mu$m$^2$). The  60~mm diameter spherical
mirror (with a focal distance of 200~mm) serves as both the camera
and the collimator providing in the Littrow combination with a
600~grooves/mm grating (maximum concentration at 6500~\AA, the size
of the ruled domain $30\times30$~mm$^2$) a reciprocal linear
dispersion of  80~\AA/mm. The single-pixel 13-$\mu$-wide entrance
slit corresponds to  $\arcsec{0.5}$. The device is designed for
spectroscopy in the H$\alpha$ region; the main problem was to
accurately center the star along the slit height.
Denby et.~al.~\cite{denby1986instr} developed a compact spectrograph for a
$D=0.6$~m telescope with the following parameters: a folded lenses
collimator with $F=485$~mm ($1:9$), a 300~grooves/mm grating, an
$F=85$~mm ($1:2$) camera with a dispersion of 4~\AA{} per
13-micron pixel. The guide view is achieved through a  $F=75$~mm
($1:3.5$) lens.

The main difficulty during the first years of working with 1-D and
2-D CCD arrays was due to the low transfer efficiency, which
resulted in residual effects in sky background line subtraction.
One of the methods used to address this problem was by
preillumination of the detector surface, which increased the
readout noise. In the case of low relative contribution from
readout noise the  $S/N$ ratio is proportional to the square root
of the number of counts, $\sqrt{n}$. Or, simply, the principal
characteristic of the detector in the case of low and high $S/N$
is readout noise and quantum efficiency, respectively. To measure
integrated colors of galaxies, Rakos et.~al.~\cite{rakos1990pasp} developed an
ultra-low-resolution spectrophotometer (140~\AA{} in the
\mbox{3200--7600~\AA) interval.} Two (``object'' and
``background'') focal lens reducers ($1:15$ to $1:2$) feeding
fiber bundles are mounted at the device entrance. The bundle
outputs are located at the focus of the 108-mm diameter ($1:2$)
225~grooves/mm concave holographic grating providing a dispersion of
200~\AA/mm. The spectrum is recorded by a CCD and the device is
equipped with an offset guide. On $D=1.1$ and $1.3$~m telescopes
the spectrum of a bright galaxy ($m_V=12^{\rm m}$) with a
signal-to-noise ratio of up to $S/N=10$ could be acquired in four
10-minute long exposures during full Moon.

Mass redshift measurement, which with the introduction of photon
counters came to be used on moderate-diameter telescopes, was
further supported by the construction of an efficient Cassegrain
spectrograph of the  \mbox{1.5-m} telescope
\cite{fabrikant1998pasp}. With  a collimated beam diameter of
100~mm, a slit width of $\arcsec{1.5}$, and a 300~grooves/mm grating the
spectral resolution is 3~\AA, and the simultaneously registered
wavelength interval spans 4000~\AA. The slit length is
$\arcmin{3}$, its efficiency 26\%, and the number of spectra
acquired in a year exceeded 10\,000, i.e., the spectrograph
replaced the famous  Z machine~\cite{latham1982instr} with a
$1^{\rm m}$ gain in terms of limiting magnitude. The main
shortcoming of suspended Cassegrain spectrographs is their
mechanical flexibility, which in each particular case makes it
difficult to construct a system of radial velocities measured with
the given device. Munari and Lattanzi~\cite{munari1992pasp} studied flextures of two
Cassegrain spectrographs of Asiago Observatory. Shifts were found
that corresponded to an error of 10--40~km\,s$^{-1}$, whereas the
error of cross-correlation technique was just 0.8~km\,s$^{-1}$. A
numerical model was constructed, which made it possible to improve
the accuracy of radial-velocity measurements, with night-to-night
errors reduced by one order of magnitude, i.e., down to
1--3~km\,s$^{-1}$.

ANS consortium developed three models of suspended spectrographs
for \mbox{$D=0.61$~m,} $0.70$~m, and $0.84$~m telescopes
\cite{munari2014multimode}. Two of these models (Mark\,II and
Mark\,III) can be transformed into cross-dispersion devices
without unmounting the spectrograph from the Cassegrain focus.

Panchuk et.~al.~\cite{panchuk2015podvesnoj} reported the development of a
cross-dispersion spectrograph for \mbox{0.6--1.0~m}-diameter
telescopes. The two-pixel spectral resolution is $R\sim40\,000$.
The simultaneously recorded wavelength interval (350--800~nm) is
determined by the spectral response curve of the detector and not
by the format of the frame. There is complete overlap of
neighboring spectral orders. A mirror collimator ($d_{\rm
{coll}}=50$~mm) is used with a lens camera
\mbox{($F_{\rm{cam}}=300$~mm, $1:3.5$).} The spectrograph slit
width ratio is 2.2 with no magnification at the echelle
($\alpha=\beta$). The angular width of the normal slit matched to
the resolution element is $\arcsec{1.6}$ and $\arcsec{1}$ on  0.6~ and
1.0~m telescopes, respectively. The mathematical apparatus for
extracting one-dimensional spectra from two-dimensional echelle
images used for the Nasmyth Echelle Spectrograph (NES) of the
6-telescope of the Special Astrophysical Observatory of the
Russian Academy of Sciences~\cite{panchuk2017hires} was refined
to be operated with a prism as the cross-dispersion element.

The low-resolution spectrograph ADAM developed for  AZT-33IK
telescope (\mbox {$D=1.6$}~m)~\cite{afanasiev2016adam} underwent
initial tests on the \mbox{Zeiss-1000} ($D=1$~m) telescope of the
Special Astrophysical Observatory of the Russian Academy of
Sciences. A lens converter was used for matching with the 1-m
mirror. When observed with  $\arcsec{1.5}$ wide slit $S/N=[6; 7]$
could be achieved for the spectrum of a $m_R=20^{\rm m}$ object
with a 30-minute long exposure. Observations with a 1.6-m
telescope produced spectra of such an object with a
signal-to-noise ratio of $S/N=[10; 15]$. The spectrograph operates
in the 3600--10\,000~\AA{} wavelength interval with a spectral
resolution of 6--15~\AA{} ($R=[1320; 270]$). As a dispersion
element three volume-phased holographic gratings (VPHG) are used,
which are mounted on a turret: one 300~grooves/mm grating (for the
entire working wavelength range) and two 600~grooves/mm gratings
\mbox{(3588--7251~\AA{}} and 6430--10\,031~\AA).

\subsubsection{Single order fiber-fed spectrographs}   
The first successful experiment involving fiber matching of a
spectrograph and a $D=0.91$~m telescope \cite{angel1979operation}
was carried out within the framework of  FLOAT project (an array
of telescopes connected with optical fibers
\cite{angel1977very}). To assess the efficiency of optical fiber
matching on a  $D=0.91$~m telescope the same spectrograph with an
image tube was used as in the Cassegrain focus of the $D=2.28$~m
telescope. Optical fiber was found to have appreciable loss in the
ground-based ultraviolet. Furthermore, fiber matching was also
shown to have advantages over the coud\'e focus where one has to
maintain high reflectivity of the mirrors in the optical path.

Another solution of fiber-optics spectrographs operating in low
diffraction orders worth noting is the Ebert-Fastie design and its
Newton modification~\cite{furenlid1988pasp, barry2002pasp}
developed within the framework of the promising Multi-Telescope
Telescope (MTT) concept. The primary focus of each of the nine
mirrors ($D=0.33$~m) of such  telescope\ \cite{bagnuolo1990pasp}
is connected to the spectrograph via a fiber. The effective
aperture of MTT is equivalent to that of a telescope with a \mbox
{$D=1.3$}~m diameter mirror.

\subsubsection{Fiber-fed echelle spectrographs}
The increase of the accuracy of radial-velocity measurements is
limited by the fundamental property of slit spectrographs. Because
of atmospheric dispersion positions of monochromatic star images
differ and the filling pattern of the collimator optics depends on
the orientation of the telescope due of residual
nonadjustments and mechanical flextures. Whereas the former
problem can be solved by installing an atmospheric dispersion
compensator, to address the latter, one has to feed the
spectrograph input a flux with constant angular aperture,
unchanging intensity distribution along the angle independent of
seeing and object tracking accuracy. That is why the progress of
fiber-optics technology has soon been put to use in astronomical
spectroscopy. In the case of small telescope the possibility of
mounting the spectrograph outside the telescope was of no small
importance (the sizes of Cassegrain echelle spectrographs are
comparable to those of the  $D\sim [0.5; 0.9]$~m telescopes).

One of the first attempts to achieve high position accuracy on a
fiber-fed echelle spectrograph was the study of
Kershaw and Hearnshaw~\cite{kershaw1989southern} who used a suspended echelle
spectrograph \cite{hearnshaw1977casseg} combined with the
MacLellan telescope ($D=1.0$~m) of Mt~John Observatory. The above
authors used a liquid-nitrogen-cooled Reticon diode array and then
a CCD as a detector. Cross-correlation techniques made it possible
to achieve an accuracy of 50~m\,s$^{-1}$ (with \mbox
{$R\sim35\,000$}) for \mbox{$m_V\le7^{\rm m}$} stars
\cite{murdoch1993search}. Long-term efforts of New Zealand
spectroscopists culminated in the construction of the large
HERCULES spectrograph \cite{hearnshaw2002exast}. The increase of
the collimated-beam diameter from 45 to 210~mm, stabilization of
conditions inside the spectrograph volume as a result of the use
of highly efficient optical coatings combined with the use of
efficient optical technologies made it possible to achieve an
accuracy of \mbox{$\sigma_{V_r}=[4; 14]$~m\,s$^{-1}$}  with a peak
efficiency of 18\% under $\arcsec{1}$ seeing conditions on a
\mbox{$D=1$~m} telescope. The collimated beam overfills the
echelle (R2, \mbox{$400\times200$~mm)}, and a ``folded Schmidt''
type camera \mbox{($F_{\rm {cam}}=973$~mm}, \mbox{$D_{\rm
{cam}}=525$~mm)} is used. The spectral resolution is determined by
the fiber-and-slit combination  \mbox{($R=41\,000$, $70\,000$,
$82\,000$)}.

In the Flash spectrograph ($d_{\rm
{coll}}=80$~mm) tested on the $D=0.75$~m  Heidelberg telescope
the echelle (R2, 31.6~grooves/mm) operates in the principal plane,
and this layout reduces the projected diameter of the fiber core
on the CCD \cite{mandel1988impact}. The spectrograph was moved to
the $D=0.5$~m telescope of European Southern Observatory
\cite{wolf1993messenger} to monitor hot southern-sky supergiants.
The Heidelberg Extended Range Optical Spectrograph
(HEROS)~\footnote{\protect\url{http://www.lsw.uni-heidelberg.de/projects/instrumentation/Heros/}}
with the ``blue'' and ``red'' branches was constructed.
Splitting the dispersed beam into two branches (3450--5600~\AA{}
and 5800--8650~\AA) made it possible to optimize the echelle order
packing density in each branch (by using cross-dispersion gratings
with different grooves densities). In each branch CCDs were used that
were optimized in terms of format and quantum efficiency. The
spectrograph was mounted at European Southern Observatory, where
it was also used in the program of the monitoring of hot
southern-sky supergiants over 120 days during six months.
Observations with HEROS spectrograph were also conducted on the
$D=0.9$~m Dutch telescope transferred to ESO from Southern
Africa~\cite{lub1979messenger}. The accuracy of radial-velocity
measurements was \mbox {$\sigma_{V_r} < 1$}~km\,s$^{-1}$, which
was considered sufficient for the program of the monitoring of hot
stars. HEROS was later moved to the $D=2$~m telescope of Ond\v
rejov Observatory~\cite{slechta2002pasi}. In 2005 the \mbox
{$D=0.5$}~m ESO telescope was moved to the University Observatory
near Santiago (Chile), where it is now operated with  PUCHEROS
fiber-fed spectrograph~\cite{vanzi2012pucheros}. The parameters
of the spectrograph are: fiber-core diameter 0.05~mm,
\mbox{$d_{\rm {coll}}=33$~mm,} 44.4~grooves/mm echelle, blaze angle
$\arcdeg{70}$, cross-dispersion unit consisting of two $\arcdeg{48}$
prisms, a \mbox{($F=355$~mm)} lens doublet camera with a meniscus
field corrector, $R=17\,800$.

Baudrand and Bohm~\cite{baudrand1992aap} developed a fiber-fed spectrograph
designed for observations within the framework of MUSICOS program.
This inexpensive device for  $D\sim2$~m telescopes distributed
along the longitude was also suitable for observations on small
telescopes.

Some of the suspended echelle spectrographs pointed out in
Table~\ref{tab_2_panchuk} were later equipped with modern
detectors combined with fiber feed. Thus one of the copies of the
spectrograph based on the successful design of the Harvard College
spectrograph \cite{latham1977inastro} was mounted on the
$D=1.0$~m telescope of Toledo University Observatory and has been
operated for several decades in the fiber-fed mode with a CCD
(see, e.g.,~Morrison~\cite{morrison1995bull}).

There are also well-known echelle spectrographs that were not
manufactured at observatories. REOSC company developed a universal
spectrograph operated both in the single-order version ($R=1000$)
and in the cross-dispersion layout ($R=21\,000$, 3800--8000~\AA\
wavelength interval). One of such spectrographs, which was earlier
operated in the $F/15$ Cassegrain focus of the  $D=0.91$~m
telescope mounted on a slope of Etna volcano is currently operated
in the fiber-fed
mode~\footnote{\protect\url{http://w3c.ct.astro.it/sln/strumenti.html}}
\cite{marino1998ibvs}. Note that this assembly provides a
radial-velocity measurement accuracy of
$\sigma_{V_r}<0.3$~km\,s$^{-1}$, which is sufficient for most of
the tasks involving spectroscopic monitoring of variable stars. It
has a commercial CANON EF300 lens installed instead of the Schmidt
camera, allowing the entire collimated beam to be used in the
fiber-fed mode.

The fiber-fed   CORALIE echelle
spectrograph~\cite{queloz2000aa}, which is an improved twin of
ELODIE spectrograph~\cite{baranne1996aass}, is installed at ESO
in the Nasmyth focus of  Euler Swiss telescope ($D=1.2$~m) and
serves to search for extrasolar planets via measuring the Doppler
shifts of the spectrum of the central star. The combination of
Euler telescope with  CORALIE spectrograph is fully automated~\cite{weber2000fully}.

In addition to telescopes dedicated for a specific class of tasks, robotic
telescope facilities, e.g., STELLA \cite{strassmeier2001stella,
strassmeier2004stella}, are developed. The parameters of this spectroscopic and
photometric facility are optimized for the study of the structure and dynamics
of the activity at the surfaces of stellar photospheres. The facility
incorporates STELLA-I ($D=1.2$~m)~--- the first fully robotic telescope equipped
with SES high-resolution echelle spectrograph ($d=130$~mm, 390--860~nm, R2,
31~grooves/mm). It is operated with two fibers, providing a resolution of
$R=50\,000$ and $R=25\,000$ for the entrance aperture of~$\arcsec{1.7}$
and~$\arcsec{3.4}$, respectively.

For Mercator telescope ($D=1.2$~m)  HERMES spectrograph was
developed \cite{raskin2008spie}, which is highlighted by its
record combination of quantum efficiency  (25\% at the peak) and
spectroscopic resolution ($R=85\,000$).

Most of the fiber-fed echelle spectrographs of the 1990s are
used with  \mbox{$1.5$--$3.6$~m} telescopes
\cite{panchuk2011high}, but it should be borne in mind that
methods of radial-velocity measurements based on echelle spectra
with minimum $S/N<1$ were also developed~\cite{queloz1995new}.
Thus radial-velocity measurements with fiber-fed echelle
spectrographs can also be performed on  \mbox{$D=[0.5; 0.9]$~m
telescopes.}

For Russian meter-class telescopes  EFES spectrograph
\cite{panchuk2015design} was developed whose prototype
\cite{panchuk2011high} has been operated since 2010 on the
$D=1.2$~m telescope of Kourovka Observatory of Ural Federal
University~\cite{punanova2013ika}.

An array of  $D=0.7$~m telescopes \cite{swift2015jast} has
been designed for the spectroscopy of candidate exoplanet system
stars. The facility also performs photometric observations and
that is why central screening covers almost half the diameter of
the primary mirror. The promising KiwiSpec model
\cite{gibson2012kiwispec, barnes2012kiwispec} based on the
asymmetric white pupil scheme (\mbox {$R=80\,000$},
5000--6300~\AA) serves as a spectrograph. Such a limited
wavelength range is caused by the need to display each of the
26 spectral orders six times (four ``science'' and two calibrating
spectra) on the $2{\rm K}\times2{\rm K}$ CCD. Calibration can be
performed not only in the classical way (via a hollow-cathode
lamp), but also via a Fabry--Perot standard.

Besides the errors mentioned in~\cite{panchuk2015doppler}, the
main factor that limits the positional and photometric accuracy of
high-resolution fiber-fed spectrographs is modal noise, which
results in nonidentical signal correction at the calibration
stage. Simply speaking, fiber input illumination variations and
flextures of the multimode optical fiber in the process of signal
accumulation prevent achieving high $S/N$ whatever the calibration
method. For example, it was shown in laboratory experiments with
\mbox{$R\sim150\,000$}~\cite{baudrant2001modal} that
\mbox{$S/N\sim500$} can only be achieved in simplified arithmetic
estimates. A comprehensive solution is provided by the transition
to a single-mode fiber where the aperture of the emerging beam is
always Gaussian. The size of an operating (diffraction-limited)
spectrograph will also be smaller. For example, in the case of the
study of Schwab et.~al.~\cite{schwab2012single} the spectrograph with R4 echelle
and  $d_{\rm{coll}}=25$~mm provides a spectral resolution of
$R\sim100\,000$. However, the core diameter of a single-mode fiber
is several times (or by one order of magnitude) smaller than that
of a multimode fiber and the solution of the problem of matching
at the input to the optical fiber depends on the adaptive optics
tools employed.

\subsection{Spectropolarimeters}
Examples of spectropolarimetric methods including the cases where they were used
on moderate-diameter telescopes can be found in Klochkova
et.~al.~\cite{klochkova2005bull}. Fine spectropolarimetric effects, which show
up in the profiles of spectral lines, are apparent only at high $S/N$, i.e.,
they can be studied with large telescopes. In the case of small telescopes
low-resolution spectropolarimetry is an option when effects in the continuum,
e.g., specificities of stellar and circumstellar polarization, can be studied.
For this reason here we mention one of the devices that  occupies a ``niche''
between narrow-band photopolarimeters and moderate-resolution
spectropolarimeters,~--- the  HBS spectropolarimeter  \cite{kawabata1999pasp}
developed for the \mbox {$D=0.9$}~m telescope. In front of the lens diffraction
spectrograph a classical polarimeter is installed with phase-shifting plates and
a Wollaston prism operating in the converging beam. The spectroscopic resolution
is 40--200. Let us point out some of the tasks that can be solved with such a
facility. Wavelength dependence of interstellar polarization differs from that
of circumstellar polarization~\cite{serkovski1975apj}. The task of their
separation is based in the relation between the wavelength of maximum
polarization and the total-to-selective extinction ratio ($R=5.5\lambda$). If
extra local effects show in the vicinity of stars then $R>3$, i.e.,
$\lambda_{\rm max}>0.55$. Hence low-resolution mass spectropolarimetry of stars
can be used to find circumstellar envelopes. Another kind of tasks involves
monitoring of circumstellar polarization. It is known that variable broadband
polarization in  T\,Tau and Be stars can amount to  10\% and 1\%, respectively.
In the case of observations with medium spectral resolution polarization effects
can be referred to separate spectral fragments or features. For example,  Be
stars have smaller polarization degree of the emission spectrum
\cite{panchuk2017izvvuz}. The proper variable polarization of M-type
supergiants (up to~2\%) caused by giant convective cells is also better to study
with low spectral resolution.

The famous spectropolarimeter of Washington State University based
on the  Boller and Chivens spectrograph ($R\sim800$) was used on the $D=0.91$
and $1.0$~m telescopes of  Pine Bluff and Ritter Observatories,
respectively \cite{davidson2014hpol}.

\subsection{Interference spectroscopic devices}
The application of interference spectrometers in astronomy is
based on two factors. First, Fellgett~\cite{fellgett1958jpr} pointed out
that if the proper noise of the detector dominates then a two-beam
interferometer is more efficient than a monochromator. Second,
Jacquinot~\cite{jacquinot1957josa} considered multibeam interferometer with
a Fabry--Perot etalon as a high-resolution monochromator. In the
era of single-channel detectors the main task was to separate one
of the orders of the etalon.

\subsubsection{Scanning Fabry--Perot interferometer with a photomultiplier}
Geake and Wilcock~\cite{geake1957mnras} implemented the method where lines in
stellar spectra were studied by tilting the Fabry--Perot interferometer (FPI).
They used a two-prism monochromator (25~\AA/mm at~H$\gamma$) mounted in the
Newton focus of the 120-cm telescope of Asiago Observatory. The beam emerging
from the monochromator was collimated \mbox{($F_{\rm {coll}}=90$~mm)}  and
passed through the etalon. The role of the monochromator reduces to suppressing
all transmission bands of the etalon except one. The wavelength transmitted by
the etalon must vary at a constant rate and to this end the etalon was tilted
via a camshaft mechanism (the cosine of the tilt angle varied linearly with
scanning time). The wavelength at the monochromator output simultaneously varied
so that the transmission band of the etalon would remain at the center of the
monochromator transmission band. The 0.09-mm thick separator of the plates of
the Fabry--Perot standard ensured 11~\AA\ gap between the orders of the
standard, and the transmission of the standard was equal to 60\%. A
factor-of-three gain was achieved in this first experiment compared to a
monochromator without a etalon.

\subsubsection{Polarization interferometers}
The use of interference spectrometers to study point objects with
small telescopes is not restricted to methods involving crossing
with prismatic or diffraction devices. The idea of a polarization
interferometer incorporating two crystal wedges placed between two
polarizers and moving in opposite directions toward each other was
first implemented by Bakhshiev in 1956 \cite{tarasov1968spectr}.
Mertz~\cite{mertz1958jpl} then tested a
similar scheme of multichannel stellar spectrometer with a
single-channel detector, which was also based on the interference
of rays with different polarization. First, the only main element
of the scheme was a Soleil compensator placed between the two
polarizers. The first polarizer consists of a Wollaston prism and
a half-wave plate covering only one image and turning its
polarization plane. This solution makes it possible to use both
polarizations. It involves measuring the difference between the
halfwave shifted systems of bands. Scintillation proved to be the
main source of noise. Then a potassium dihydrophosphate plate was
added to the scheme. When subject to a longitudinal electric
field, this plate changes the path difference as a result of
double refraction. Lock-in detection was performed with a
frequency of 3~kHz. The polarization interferometer was tested in
the Cassegrain focus ($1:18$) focus of the $D=0.6~$m telescope.
With low-resolution observations it was planned to use
interferograms directly for the classification of spectra
(observations were made in the pre-computer era). Serkowski
\cite{serkowski1972pasp} proposed a method for studying the
distribution of radial velocities in extended objects with
emission-line spectra. The line studied is separated by an
interference filter and then light passes through the
polarization interferometer and is then registered by a field
detector. Exposures are made for different turn angles of the
phase-shifting plate of the interferometer. The polarization
position angle at each point of the nebula can be calibrated in
terms of radial velocities. The method provides higher angular
resolution compared to FPI.

\subsubsection{Multichannel spectrographs with a FPI}
We illustrate the use of multichannel systems in interferometric
devices with two examples. Serkowski~\cite{serkowski1978high} mounted a FPI
at the entrance of the echelle spectrograph in the Cassegrain
focus ($1:13.5$) of a \mbox{1.54-m} telescope. The light in the
4100--4400~\AA{} part of the spectrum was separated by a
preliminary dispersion grism unit (70~\AA/mm), and then arrived to
the spectrometer where eight echelle orders were registered by a
brightness amplifier and a $342\times42$ diode array (Digicon). A
quartz plate is placed behind the entrance diaphragm. This plate
can take two positions in terms of tilt angle. These tilt angles
ensure shifting the FPI order on the diode array to the location
of the neighboring order. To compensate the wavelength dependence
of the free spectral interval of the FPI the thickness of the
quartz plate varies along the grism dispersion direction. The FPI
transmission bands have the width of 0.06~\AA, and the separation
between neighboring orders is 0.62~\AA{} at 4250~\AA, which for
the reciprocal linear dispersion 3.4~\AA/mm corresponds to
five-pixel separation between neighboring FPI orders whose images
have two-pixel-sized diameters. The vacuum camera with the FPI is
tilted by a precision device within $\pm\arcdeg{1}$, and the total
observing run consists of 20 exposures with recording the points
of the spectrum 0.03~\AA\ apart. Interferometer tilt angle and
wavelength calibration is performed by recording the comparison
spectrum of a hollow-cathode lamp and photodiode registration of
the pair of \mbox{He-Xe} laser beams separated by $\arcdeg{6}$. The
observing technique prevents any effect of the atmospheric
transparency, photocathode sensitivity, and seeing variations. The
accuracy achieved with 20 FPI tilt positions for a sixth-magnitude
star corresponds to a radial-velocity error of 10~m\,s$^{-1}$.

The second attempt to use interferometry technique for measuring
the Doppler shift was also made by the staff of the University of
Arizona Observatory. McMillan et.~al.~\cite{mcmillan1993aj} developed a fiber-fed
echelle spectrograph with a CCD where the FPI operated in the
inner installation, i.e., it was placed in the collimated beam. A
total of 350 FPI orders were recorded simultaneously in the
spectral orders of the echelle spectrum covering the
\mbox{4250--4600~\AA} wavelength interval. The width of a FPI
order at 4300~\AA{} is 47~m\AA{} and the neighboring orders are
0.64~\AA\ apart. Doppler shifts in the stellar spectrum change the
order intensity ratios. For the sake of simplicity, only velocity
variations were recorded, i.e., the device operated as an
accelerometer. The argon-glow lamp calibration provided an
accuracy reaching two one-hundred millionth, which corresponds to
a velocity error of $\pm6$~m\,s$^{-1}$. Instrumental variations
within $\pm27$~m\,s$^{-1}$ were found on a time scale of several
months. The device was used on a $D=0.9$~m telescope.

\subsubsection{Spectrographs with an external interferometer}
To analyze spectra of extended sources,
Panchuk~\cite{panchuk2000prepsao144} proposed the method of twice crossed
dispersion whose idea consists in measuring a single-sided fringe
system of the FPI mounted in front of the echelle spectrograph.
However, a high-Q-factor Fabry--Perot interferometer does not
produce sine waves like the Michelson interferometer, and hence
the advantages of Fourier analysis cannot be used for precision
determination of the phase shift. Furthermore, the FPI transmits
less light than the Michelson interferometer at the peaks of the
instrumental function. Erskine~\cite{erskine2003pasp} proposed a scheme
where the Michelson interferometer is crossed with a diffraction
spectrograph, and demonstrated its efficiency in measuring radial
velocities with an accuracy that allows recording the displacement
of the Earth--Moon barycenter (the variation amplitude is
12~m\,s$^{-1}$). Compared to heterodyne holographic spectrograph
\cite{frandsen1993aap}, the working wavelength range of the
device is broader by tens of times.

Externally dispersed interferometry (EDI) deserves a separate
consideration as a promising development. Here we only point out
that the first Doppler-based detection of an exoplanet using this
method was performed on a  $D=0.9$~m telescope and then confirmed
on large telescopes~\cite{jian2006first}. The half-amplitude of
radial-velocity variations of a $m_V=\m{8.05}$ star was
\mbox{$63.4\pm2.0$~m\,s$^{-1}$} with a period of 4.11~days. This
is how a companion with a minimum mass  $m\sin i$ equal to 0.49
Jupiter masses was discovered.

\subsection{Small-telescope spectrographs}
In this section we mention spectrographs meant for the use on
small telescopes ($D=[0.25; 0.4]$~m) with various detectors and in
various combinations with the telescope.

In Canopus Hill Observatory (University of Tasmania) a $D=0.4$~m
reflector equipped with coud\'e focus ($1:33$) came into operation
in 1977 \cite{castley1972pasa}. The parameters of the
spectrograph are: \mbox {$F_{\rm {coll}} = 330$}~cm, $d=100$~mm,
two diffraction gratings with a \mbox{$152\times102$~mm$^2$} ruled
area, 600~grooves/mm (used in the first order) and 1200~grooves/mm (in
the second order), and two cameras, $F_{\rm {cam}} = 122$ and
$182$~cm with the 76~cm camera mirror diameter. The reciprocal
linear dispersion  values were $P = 2.4$, $3.6$, $9$, and
$14$~\AA/mm. Note that the diameter of camera mirrors is greater
than the diameter of the primary mirror of the telescope, so that
the cost of the spectrograph is comparable to the cost of the
telescope. The results of photographic spectroscopy of bright
F-type supergiants can be found in Castley and Watson~\cite{castley1980aas}.
Research work at Canopus Hill were discontinued in 2013 because of
ever increasing light pollution.

A stigmatic echelle spectrograph \mbox{($R=40\,000$)} with an
image tube (electrostatic focusing, \hbox{S-25} cathode) was used to
search for water bands (8200~\AA range) in the atmosphere of
Venus~\cite{gull1974icarus}. To reduce the contribution from the
spectrum of tropospheric water vapor, observations were made from
onboard an aircraft at a height of 14.6~km during the period of
the elongation of Venus (April, 1972), when the radial-velocity
difference between the atmosphere of Venus and telluric
absorptions was maximal. A $D=0.25$~m telescope was used as
feeding optics. The parameters of the spectrograph are: Newton
collimator, \mbox{$d_{\rm {coll}}=150\,$mm}; 79\,grooves/mm,
\mbox{$\tan\theta_b = 2$} echelle with the  \mbox
{$150\times300$}~mm$^2$ ruled area, operating in the primary plane
($2\theta = \arcdeg{12}$); a 300\,grooves/mm cross-dispersion grating
with a \mbox{$200\times250$~mm$^2$ ruled ares.} In this case the
spectrograph is even more expensive that the telescope optics.

Currently, the manufacturers offer spectrographs oriented toward
both professional and amateur astronomers. These devices, which
can be mounted on  $D=[0.2; 0.4]$~m telescopes, are also used in
programs of spectroscopic monitoring of bright stars~\cite{miroshnichenko2013periastron}.

BACHES Cassegrain echelle spectrograph equipped with a
CCD~\cite{avila2007high} appears to be of commercial interest and
the parameters of the device are not fully disclosed: it has a
$1:10$ collimator; 79~grooves/mm, \mbox{$\tan\theta_b = 2$} echelle;
a cross-dispersion grating, and a lens camera. The  \mbox
{$1530\times1020$} CCD with a pixel size of $9\times9$~$\mu$m$^2$
simultaneously records 29 spectral orders in the
\mbox{3900--7500}~\AA\ wavelength interval. The  entrance slit
projection onto the detector has a size of 2.4 pixels and the
spectral resolution is $R=19\,000$. On a \mbox{$D=0.25$~m}
telescope the spectrum of a fifth-magnitude star is acquired with
$S/N = 50$ in a 900-second exposure under  $\arcsec{1.7}$ seeing
conditions. The spectrograph transmission is equal to 27\% at
5040~\AA. The quantum efficiency of the system (the atmosphere,
telescope, spectrograph, and detector) is equal to 11\%. Recall
that under the same conditions UVES VLT spectrograph has a quantum
efficiency of 17\%. Before starting serial production of the
spectrograph it was field tested on a  $D=0.5$~m telescope
\cite{kozlowski2014baches}, and short-scale radial-velocity
errors ($\sigma_{V_r}=[1.5; 1.7]$~km\,s$^{-1}$) were shown to be
determined by the flexibility of the standard adaptor and can
therefore be reduced. BACHES spectrograph is equipped with two
slits and is optimized for a \mbox{$D=0.25$~m} ($1:10$) telescope.

For a telescope of the same diameter an echelle spectrograph was
developed that is oriented toward the study of  $m_V<6^{\rm m}$
stars \cite{panchuk2015spectrograph}. Its  $1:4$ lens collimator
forms a \mbox{$d=30$~mm} beam and the device is equipped with a
75~grooves/mm, $\tan\theta_b = 2$ echelle and a $1:2$ lens camera.
The two-pixel spectral resolution is $R=16\,000$ for the normal
slit width $s=\arcsec{4.4}$. Echelle operates in the autocollimation
mode ($\alpha=\beta=\theta_b$, outside the primary plane, i.e.,
$\gamma\ne0$). In this case the dependence of energy concentration
along the order varies more steeply and at maximum it is 20--30\%
greater than in the case $\alpha>\theta_b>\beta$ and $\gamma=0$,
i.e., in the primary plane. The device's cross-dispersion unit is
a  300~grooves/mm grating operating in the first order. The latter
circumstance made it possible to make design provisions for the
change of $\gamma$ (half of the  $2\gamma$ angle between the
collimator axis and the ``echelle center--cross-dispersion grating
center'' line) to place the selected line to the maximum of the
energy concentration curve in the echelle order. For example,
varying  $\gamma$ from $\arcdeg{6}$ to $\arcdeg{8}$ changes the central
wavelength in the $m=36$ order from \mbox{$\lambda_c=6593$}~\AA{}
to $\lambda_c=6564$~\AA. In the case of such a design the optimum
choice is an R2 echelle with a line density of 37.5~grooves/mm,
where the length of the order is twice shorter and the range of
$\gamma$ is twice smaller. A Meade LXD55 telescope,
\mbox{($D=254$~mm}, $F=1016$~mm) is used as feeding optics. The
spectrograph used in the Newton focus is fixed parallel to the
telescope tube.

Eagleowloptics company (Switzerland) developed compact SQUES spectrograph
($R=20\,000$).

Suspended single-order  DADOS spectrograph is not
intended for achieving maximum spectral resolution (its grating
operates with the right angle between the incident and diffracted
beams). It is operated with two~--- 200 and 900~grooves/mm~--- gratings, $F_{\rm {coll}}=80$~mm
($1:10$),  and \mbox {$F_{\rm
{cam}}=96$}~mm. When operated with the 900~grooves/mm grating the
reciprocal dispersion and slit width are \mbox{$P=106$~\AA/mm} and
0.025~mm, respectively. When mounted on a  \mbox{$D=0.3$~m}
telescope this spectrograph acquires $S/N=50$ spectrum for a
$m_V=6^{\rm m}$ star in 20 minutes.

LHIRES\,III suspended single-order high-resolution spectrograph is
designed in accordance with autocollimation scheme: the same
doublet serves as both the collimator and the camera, $F=200$~mm.
Provision is made for the use of replaceable diffraction gratings
with line densities spanning from 150 to 2400~grooves/mm. The
spectrograph is optimized for a \mbox{$D=0.2$}~m ($1:10$)
telescope. When equipped with a 2400~grooves/mm grating and a
detector with a \mbox{9-$\mu$m}-sized pixel, the spectrograph
provides a resolution of \mbox{$R=17\,000$.} The signal-to-noise
ratio for a fifth-magnitude star observed with one-hour long
exposure is $S/N=100$. The spectrograph was used extensively in
programs of monitoring of hot stars with H$\alpha$ emission.

In the 1990s  SBIG started selling  SGS model where two
replaceable gratings operate in accordance with the Ebert scheme
with a small spherical mirror. In  DSS7 model lens optics is used
for both the camera and the collimator, $P=600$~\AA/mm. The
company equips its spectrographs with its own-manufactured
detectors, which allowed it to use digital image stabilization.

Astro Spectroscopy Instruments EU (Potsdam) developed miniature
MiniSpec spectrograph for $1:5$ and $1:10$ telescopes. Options
(combination of the slit width and grating line density) are
provided that allow the spectrograph to be operated on a detector
with 9-$\mu$m pixel size and a reciprocal linear dispersion of
\mbox{$P=[0.2; 3.3]$}~\AA/pixel.

An autocollimation spectrograph was developed for Celestron CPC
1100 telescope \mbox{($D=279$}~mm, $F=2800$~mm,
Schmidt--Cassegrain) \cite{panchuk2015autocoll}. The
collimator/camera has a mirror lens with $D=45$~mm, $F=275$~mm.
The 1800~grooves/mm grating with a $50\times50$~mm$^2$ ruled area
operates in the first diffraction order. The reciprocal linear
dispersion in the neighborhood of yellow mercury doublet is
$P\sim10$~\AA/mm. The maximum (theoretical) two-pixel (20~$\mu$m)
spectral resolution is $R=28\,900$, which corresponds to the
$\arcsec{1.4}$ slit width. The spectrograph is meant for monitoring
selected lines the spectra of bright ($m_V<6^{\rm m}$) variable
stars of various types with the resolution typical for the Main
Stellar Spectrograph the 6-m telescope of the Special
Astrophysical Observatory of the Russian Academy of Sciences
\cite{panchuk2014main}, \mbox{$R=14\,000$.}

The spectrographs for small-diameter telescopes are so far
dominated by suspended devices because of their small size and
mass. However, there also fiber-fed systems, which are usually
adopted for financial reasons. First, where there is a laboratory
spectrograph it can also be used with the telescope. For example,
the $D=0.51$~m telescope at the University of Illinois
(Springfield~\footnote{\protect\url{https://www.uis.edu/}}) is equipped
with SE200 Echelette spectrograph ($R\sim20\,000$), manufactured
by  Catalina Scientific Instruments for laboratory works (it has
fiber-fed input). Its dispersing unit (echelle and prism) operates
in accordance with the Ebert scheme ($F=200$~mm, $1:10$). The
telescope is also used with an Optomechanics 10C spectrograph
(developed by Optomechanics Research of Vail, AZ.) whose layout
includes an $F=225$~mm ($1:9$) spherical mirror collimator,
replaceable gratings (with line densities spanning from 300 to
1200~grooves/mm), and an $F=135$~mm ($1:2.8$) camera. Second, a
commercial fiber-fed spectrograph is easier to adapt for small
telescopes given the great variety of their mounting types and
size restrictions. eShel fiber-fed spectrograph
\cite{thizy2011spectrographs} is based on the cross-dispersion
scheme. Its $F_{\rm {coll}}=125$~ ($1:5$) mirror collimator is fed
through an optical fiber (0.05~mm). The collimated  $d_{\rm
{coll}}=25$~mm beam is directed to the  R2 echelle whose
cross-dispersion unit is a prism. The camera has an $F_{\rm
{cam}}=85$~mm ($1:1.8$) lens. The $13\times9$~mm$^2$ detector
records only the visible part of the spectrum (4500--7000~\AA),
\mbox{$R=10\,000$.} Under~$\arcsec{3}$ seeing conditions a
\mbox{$D=0.2$~m} ($1:5.9$) telescope achieves  $S/N=100$ during
one-hour exposure of an $m_V=\m{7.1}$ star.

A fiber-fed single-order Czerny--Turner spectrograph was used on a
$D=0.4$~m telescope to study $\tau$\,Bo\"otis system
\cite{kaye2006high}. The device has the following
parameters~--- $F_{\rm {coll}}=762$~mm; $F_{\rm {cam}}=240$~mm;  1800~grooves/mm,
0.17~\AA/pix grating, and simultaneously records a  88~\AA\ wide
spectral fragment. The projected size of the fiber core is
4~pixels (0.68~\AA), \mbox{$R=7500$.} Despite such spectral
resolution, which is by no means optimum for the discovery of
exoplanets, observations made with this device confirmed the
parameters of the system distinguished by its large
radial-velocity amplitude ($K=471\pm10$~m/s). The position
of the star image was corrected eight times a second using
standard SBIG tools.

\section{Prospects}
An analysis of various technical solutions and personal practical
experience allow us to highlight some promising solutions in the
technique and organization of spectroscopy on  medium- and
small-diameter telescopes.

First of all, this is the further specialization of instruments. It goes without
saying that equipping a medium or small telescope with one instrument results in
significant economy in terms of the maintenance costs of the
telescope--spectrograph system and facilitate the transition to the remote
control mode. The review~\cite{panchuk2020issledovanie}  proposed to develop a
dedicated spectroscopic telescope with $D\sim1.2$~m.

In fiber-fed spectroscopy, lens cameras are preferred, and Schmidt
systems are underestimated due to losses in the case of center
shielding. However, the cameras of efficient  spectrographs~---  SOPHIE~\cite{perruchot2008sophie},  
STELLA and HERCULES~--- are made according to the ``folded Schmidt'' scheme, 
and the
losses on central shielding are compensated for by other
advantages (overfilling of the echelle, achromaticity of the
camera, reduced vignetting, and reduced cost for a given $d_{\rm
{coll}}$. Recall that the replacement of the spectrograph with an
ELODIE lens camera by SOPHIE has increased the quantum efficiency
of the system tenfold.

It is necessary to develop a combination of different functions on
one optical element. For example, the asphericity of reflective
gratings makes it possible to build mirror schemes whose range is
limited only by the parameters of the detector and optical
coatings. In echelle systems, an aspherical reflective grating can
operate as a cross-dispersion element. In the EMILIE fiber-fed
echelle spectrograph, one of the surfaces of the double-path prism
is made aspherical \cite{bouchy1999iaucoll}.

A dedicated spectroscopic telescope does not necessarily need to
have a large field of good images. In that case it is enough to
use a light-collecting telescope (similar to the Dutch Light
Collector \cite{lub1979messenger}). A segmented mirror is planned
to be used to correct aberrations in the spectroscopic telescope
of the Special Astrophysical Observatory of the Russian Academy of
Sciences.

It seems to a promising solution for medium and small telescopes
with  quality optics to use a diffraction-limited spectrograph
with a single- mode optical fiber. However, it should be assessed
in which tasks addressable with such telescopes, the modal noise
is a limiting factor.

Medium-resolution external-postdispersion (EDI) spectrograph
schemes are used both with fiber-fed optics  and in a suspended
version.

In the class of suspended systems:
\begin{list}{}{
\setlength\leftmargin{3mm} \setlength\topsep{2mm}
\setlength\parsep{0mm} \setlength\itemsep{2mm} }

\item 1.  Mirror schemes with low slit-width ratios and
astigmatism compensation are undeservedly forgotten \cite{schroeder1967echelle}. With then
increasing CCD format, there will be a return to these schemes on
$D\sim1$~m telescopes. For operation in the ground-based
ultraviolet, a purely mirror system with an image cutter can be
built.

\item 2. The scheme of the spectrograph using  an echelle with a
variable line density \cite{nagulin1980echelle, gerasimov1970ois}
has not become popular among the astronomers. Based on this
solution, a compact high-resolution suspended spectrograph with
maximum efficiency can be built for a medium-diameter telescope.

\item 3. The designs of the Mark series suspended slit
spectrographs proved to be a successful solution~\cite{munari2014multimode}.
\end{list}

A common problem is the need to reduce the contribution of
scattered light and ghost images. A new spectroscopic device,
which prepares the comparison spectrum, is proposed for the
two fiber spectrograph scheme \cite{panchuk2018kalibrovka}.

\section*{Conclusions}
We review the main types of spectroscopic equipment for  small-
and moderate-diameter telescopes indicating the principal
parameters of the selected design solutions. We also provide a
list of references to allow a more in-depth understanding of the
problem.

Some prospects for the development of this
instrumental and methodological direction are assessed.

In the era of the construction of large telescopes, interest in
instrumental equipment for  small~--- (less than 0.4~m) and
moderate~--- (0.4--1.2 m) diameter telescopes may seem irrelevant.
However, even a superficial assessment of the capabilities of
modern instruments of the diameters just mentioned indicates
unremitting attention to their equipment. The efficiency of these
tools even increases as some of the small diameter tools move into
the single-program category.

Small telescopes play a crucial role in the practical training of
young astronomers. The technological gap that is observed in our
country between the equipment of professional and training
telescopes seriously affects the level of training of astronomers
and physicists at universities. The aim of this publication is to
review the spectroscopic equipment of small- and moderate-diameter
telescopes and provide a brief description of the specificities of
the operation of this equipment or provide references to the
relevant literature. Despite the fact that some of the  techniques
and methods described here look archaic, one should know their
strengths and weaknesses, which in the era of highly efficient
multichannel detectors can lead to the rebirth of certain methods.
Hence the published information may prove to be useful when
reequipping domestic telescopes with new astrophysical equipment.

We believe that the instrumentation of such telescopes has large
reserves, including a combination of well-known solutions with new
technological capabilities.

\section*{Acknowledges}
V.~E.~Panchuk acknowledges support from the Government of the
Russian Federation and the Ministry of Higher Education and
Science of the Russian Federation [grant 075-15-2020-780
(No.~13.1902.21.0039)]. V.~G.~Klochkova acknowledges the support
from the Russian Science Foundation (project No.~20-19-00597).
Observations with the telescopes of the Special Astrophysical
Observatory of the Russian Academy of Sciences are supported
financially by the Ministry of Science and Higher Education of the
Russian Federation (including contract No. 05.619.21.0016, unique
project identifier  RFMEFI61919X0016).


\bibliography{Panchuk_en}

\begin{thebibliography}{199}
\providecommand{\natexlab}[1]{#1}
\providecommand{\url}[1]{\texttt{#1}}
\expandafter\ifx\csname urlstyle\endcsname\relax
  \providecommand{\doi}[1]{doi: #1}\else
  \providecommand{\doi}{doi: \begingroup \urlstyle{rm}\Url}\fi

\bibitem[Wood(1958)]{bradshaw1958present}
F.~B. Wood, editor.
\newblock \emph{The Present and Future of the Telescope of Moderate Size}.
\newblock Univ. of Pennsylvania Press (1958).

\bibitem[Scheglov(1960)]{scheglov1960present}
P.~V. Scheglov, editor.
\newblock \emph{Nastoyaschee i buduschee teleskopov umerennogo razmera}.
\newblock Moscow, Izd. Inostr. liter. (1960).

\bibitem[Hearnshaw and Cottrell(1985)]{hearnshaw1985instrumentation}
J.~B. Hearnshaw and P.~L. Cottrell, editors.
\newblock \emph{Instrumentation and research programmes for small telescopes.
  Proc. of the IAU Symp.}, volume 118.
\newblock D.~Reidel, Dordrecht (1985).

\bibitem[Panchuk et~al.(Spec. Astrophys. Obs, Nizhny Arkhyz, 2004)Panchuk,
  Emelianov, Klochkova, and Romanenko]{panchuk2004prepsao195}
V.~E. Panchuk, E.~V. Emelianov, V.~G. Klochkova, and V.~P. Romanenko.
\newblock Preprint No.~195, SAO RAS (Spec. Astrophys. Obs, Nizhny Arkhyz,
  2004).

\bibitem[Richardson(1968)]{richardson1968}
E.~Richardson.
\newblock \emph{J. Roy. Astron. Soc. Canada}, \textbf{62}, 313 (1968).

\bibitem[Bowen(1952)]{bowen1952thespectrographic}
I.~S. Bowen.
\newblock \emph{\apj}, \textbf{116}, 1 (1952).

\bibitem[Swift et~al.(Apr--Jun 2015)Swift, Bottom, Johnson,
  et~al.]{swift2015jast}
J.~J. Swift, M.~Bottom, J.~A. Johnson, et~al.
\newblock \emph{Journal of Astronomical Telescopes, Instruments, and Systems},
  \textbf{1} (2), 027002 (Apr--Jun 2015).

\bibitem[Warner(1986)]{warner1986instrumentation}
B.~Warner.
\newblock In J.~B. Hearnshaw and P.~L. Cottrell, editors, \emph{Instrumentation
  and research programmes for small telescopes. Proc. of the IAU Symp.}, volume
  118, page~3. Reidel, Dordrecht (1986).

\bibitem[Abt(2012)]{abt2012aj}
H.~A. Abt.
\newblock \emph{\apj}, \textbf{144} (4), 91 (2012).

\bibitem[Mayall(1937)]{mayall1937pasp49101}
N.~U. Mayall.
\newblock \emph{\pasp}, \textbf{49}, 101 (1937).

\bibitem[Babcock(1939)]{babcock1939lick}
H.~W. Babcock.
\newblock \emph{Lick Obs. Bull.}, \textbf{19} (498), 41 (1939).

\bibitem[Kemp et~al.(1970)Kemp, Swedlund, Landstreet, and Angel]{kemp1970apj}
J.~C. Kemp, J.~B. Swedlund, J.~Landstreet, and J.~R.~P. Angel.
\newblock \emph{\apj}, \textbf{161}, 77 (1970).

\bibitem[Angel and Landstreet(1971)]{angel1971apj}
J.~R.~P. Angel and J.~Landstreet.
\newblock \emph{\apj}, \textbf{165}, 171 (1971).

\bibitem[Griffin(1967)]{griffin1967apj}
R.~F. Griffin.
\newblock \emph{\apj}, \textbf{148}, 465 (1967).

\bibitem[Fletcher et~al.(1982)Fletcher, Harris, McClure, and
  Scarfe]{fletcher1982pasp}
J.~M. Fletcher, H.~C. Harris, R.~D. McClure, and C.~D. Scarfe.
\newblock \emph{\pasp}, \textbf{94}, 1017 (1982).

\bibitem[Mayor(1985)]{mayor1986stellar}
M.~Mayor.
\newblock In A.~G.~D. Philip and D.~W. Latham, editors, \emph{Stellar radial
  velocities. Proc. of IAU Coll.}, volume~88, pages 35--48. Schenectady:
  L.~Davis Press (1985).

\bibitem[Kurtz(1982)]{kurtz1982mnras}
D.~W. Kurtz.
\newblock \emph{\mnras}, \textbf{200}, 208 (1982).

\bibitem[Hiltner(1949)]{hiltner1949science}
W.~A. Hiltner.
\newblock \emph{Science}, \textbf{109}, 165 (1949).

\bibitem[Hall(1949)]{hall1949aj}
J.~S. Hall.
\newblock \emph{\aj}, \textbf{54} (1179), 187 (1949).

\bibitem[Hiltner(1951)]{hiltner1951apj}
W.~A. Hiltner.
\newblock \emph{\apj}, \textbf{114}, 241 (1951).

\bibitem[Pickering(1891)]{pickering1891annals}
E.~C. Pickering.
\newblock \emph{Annals of the Astron. Obs. Of Harvard College}, \textbf{26}, 1
  (1891).

\bibitem[Arnulf et~al.(1936)Arnulf, Barbier, Chalonge, and
  Canavaggia]{arnulf1936journal}
A.~Arnulf, D.~Barbier, D.~Chalonge, and R.~Canavaggia.
\newblock \emph{Journal des Observateurs}, \textbf{19} (9), 149--184 (1936).

\bibitem[Struve(1937)]{struve1937apj}
O.~Struve.
\newblock \emph{\apj}, \textbf{86}, 613 (1937).

\bibitem[Struve et~al.(1938)Struve, Van~Biesbroeck, and Elvey]{struve1938apj}
O.~Struve, G.~Van~Biesbroeck, and C.~T. Elvey.
\newblock \emph{\apj}, \textbf{87}, 559 (1938).

\bibitem[Pikelner(1954)]{pikelner1954ika}
S.~B. Pikelner.
\newblock \emph{Izv. Krymskoj Astrophyz. Obs.}, \textbf{11}, 8 (1954).

\bibitem[Mirzoyan(1955)]{mirzoyan1955photometric}
L.~V. Mirzoyan.
\newblock \emph{Communications of the Byurakan Astrophysical Observatory
  (ComBAO)}, \textbf{16}, 41--52 (1955).

\bibitem[Pickering(1896)]{pickering1896an}
E.~Pickering.
\newblock \emph{Astr. Nachr.}, \textbf{142}, 105 (1896).

\bibitem[Fehrenbach(1947)]{fehrenbach1947cnd}
C.~Fehrenbach.
\newblock \emph{Ann d'Astrophys.}, \textbf{10}, 257 (1947).

\bibitem[Gieseking(1979{\natexlab{a}})]{gieseking1979st}
F.~Gieseking.
\newblock \emph{Sky and Telescope}, \textbf{57}, 142 (1979{\natexlab{a}}).

\bibitem[Gieseking(1979{\natexlab{b}})]{gieseking1979tm}
F.~Gieseking.
\newblock \emph{The Messenger}, \textbf{17}, 29--32 (1979{\natexlab{b}}).

\bibitem[Fehrenbach and Burnage(1981)]{fehrenbach1981aaa}
C.~Fehrenbach and R.~Burnage.
\newblock \emph{\aas}, \textbf{43}, 297--306 (1981).

\bibitem[Morgan et~al.(1943)Morgan, Keenan, and Kellman]{morgan1943atlas}
W.~W. Morgan, P.~C. Keenan, and E.~Kellman.
\newblock \emph{An atlas of stellar spectra, with an outline of spectral
  classification}.
\newblock Chicago, Ill., The University of Chicago press (1943).

\bibitem[Campbell(1898)]{campbell1898mills}
W.~W. Campbell.
\newblock \emph{\apj}, \textbf{8}, 123--156 (1898).

\bibitem[Mayall(1936)]{mayall1936pasp}
N.~U. Mayall.
\newblock \emph{\pasp}, \textbf{48} (281), 14 (1936).

\bibitem[Titus and Morgan(1940)]{titus1940apj}
J.~Titus and W.~W. Morgan.
\newblock \emph{\apj}, \textbf{92}, 256 (1940).

\bibitem[Albitzky and Shajn(1932)]{albitzky1932radvel}
V.~A. Albitzky and G.~A. Shajn.
\newblock \emph{\mnras}, \textbf{92}, 771 (1932).

\bibitem[Kopylov(1954)]{kopylov1954ika}
I.~M. Kopylov.
\newblock \emph{Izv. Krymskoj Astrophyz. Obs.}, \textbf{11}, 44 (1954).

\bibitem[Hubert-Delplace and Hubert(1979)]{hubert1979atlas}
A.-M. Hubert-Delplace and H.~Hubert.
\newblock \emph{An atlas of Be stars}.
\newblock Paris-Meudon: Observatory (1979).

\bibitem[Baillet et~al.(1952)Baillet, Chalonge, and
  Cojan]{baillet1952recherches}
A.~Baillet, D.~Chalonge, and J.~Cojan.
\newblock \emph{Annales d'Astrophysique}, \textbf{15}, 144 (1952).

\bibitem[Wood(1935)]{wood1935phrev}
R.~W. Wood.
\newblock \emph{Phys. Rev.}, \textbf{48}, 928 (1935).

\bibitem[Harrison(1949{\natexlab{a}})]{harrison1949prod}
G.~R. Harrison.
\newblock \emph{\josa}, \textbf{39} (6), 413--426 (1949{\natexlab{a}}).

\bibitem[Gerasimov et~al.(1957{\natexlab{a}})Gerasimov, Tel'tevskij,
  Spizharskij, and Nesmelov]{gerasimov1957delit}
F.~M. Gerasimov, I.~A. Tel'tevskij, S.~N. Spizharskij, and S.~V. Nesmelov.
\newblock \emph{Opt.-mech. Prom.}, \textbf{3}, 47--54 (1957{\natexlab{a}}).

\bibitem[Gerasimov et~al.(1957{\natexlab{b}})Gerasimov, Tel'tevskij,
  Spizharskij, and Nesmelov]{gerasimov1957opyt}
F.~M. Gerasimov, I.~A. Tel'tevskij, S.~N. Spizharskij, and S.~V. Nesmelov.
\newblock \emph{Opt.-mech. Prom.}, \textbf{4}, 57--60 (1957{\natexlab{b}}).

\bibitem[Kossova et~al.(1958)Kossova, Nizhin, and Smirnova]{kossova1958opyt}
N.~F. Kossova, A.~M. Nizhin, and A.~V. Smirnova.
\newblock \emph{Opt.-mech. Prom.}, \textbf{8}, 35--39 (1958).

\bibitem[Epstein(1967)]{epstein1967pasp}
L.~Epstein.
\newblock \emph{\pasp}, \textbf{79} (467), 132 (1967).

\bibitem[G\'erard(1976)]{lemaitre1976reflective}
L.~G\'erard.
\newblock \emph{\josa}, \textbf{66} (12), 1334--1340 (1976).

\bibitem[Lemaitre(1981)]{lemaitre1981proc}
G.~Lemaitre.
\newblock In \emph{Instrumentation for Astronomy with Large Optical Telescopes.
  Proc. of the IAU Coll.}, volume~67, page 137. Zelenchukskaya (1981).

\bibitem[Lemaitre(1983)]{lemaitre1983proc}
G.~Lemaitre.
\newblock In \emph{Astronomy with Schmidt-type Telescopes. Proc. of the IAU
  Coll.}, volume~78, page 533. Asiago (1983).

\bibitem[Fehrenbach and Chun(1981)]{fehrenbach1981observations}
C.~Fehrenbach and H.~C. Chun.
\newblock \emph{\aas}, \textbf{46}, 257--261 (1981).

\bibitem[Hoag and Schroeder(1970)]{hoag1970pasp}
A.~A. Hoag and D.~J. Schroeder.
\newblock \emph{\pasp}, \textbf{82}, 1141 (1970).

\bibitem[Linnik(1963)]{linnik1963nta}
V.~P. Linnik.
\newblock In N.~N. Michelson, editor, \emph{Novaya technika v astronomii},
  volume~1, page 176. Leningrad: Nauka (1963).

\bibitem[Abt(1963)]{abt1963apjs}
H.~A. Abt.
\newblock \emph{\apjs}, \textbf{8}, 99 (1963).

\bibitem[Meinel(1963)]{meinel1963al}
A.~B. Meinel.
\newblock \emph{\aj}, \textbf{68}, 285 (1963).

\bibitem[Wilson(1956)]{wilson1956pasp}
O.~C. Wilson.
\newblock \emph{\pasp}, \textbf{68} (403), 346 (1956).

\bibitem[Rachkovskaya(2013)]{rachkovskaya2013ika}
T.~M. Rachkovskaya.
\newblock \emph{Izv. Krymskoj Astrophyz. Obs.}, \textbf{109} (2), 118--128
  (2013).

\bibitem[Chentzov(2013)]{chentzov2013ika}
E.~L. Chentzov.
\newblock \emph{Izv. Krymskoj Astrophyz. Obs.}, \textbf{109} (2), 152--155
  (2013).

\bibitem[Vitrichenko et~al.(1975)Vitrichenko, Volkov, Shanin, Shevchenko, and
  Shcherbakov]{vitrichenko1975infrared}
E.~A. Vitrichenko, I.~V. Volkov, G.~I. Shanin, V.~S. Shevchenko, and A.~G.
  Shcherbakov.
\newblock \emph{Soviet Astronomy}, \textbf{18}, 513 (1975).

\bibitem[Afanasiev and Pimonov(1981)]{afanasiev1981opyt}
V.~L. Afanasiev and A.~A. Pimonov.
\newblock \emph{Izvestiya SAO}, \textbf{13}, 76--84 (1981).

\bibitem[Bychkov et~al.(1978)Bychkov, Morozova, and Panchuk]{bychkov1978rapid}
K.~V. Bychkov, S.~M. Morozova, and V.~E. Panchuk.
\newblock \emph{Sov. Astr. Letters}, \textbf{4}, 170 (1978).

\bibitem[Morozova and Panchuk(1978)]{morozova1978some}
S.~M. Morozova and V.~E. Panchuk.
\newblock \emph{Soobshch. Spets. Astrofiz. Obs.}, \textbf{22}, 27--42 (1978).

\bibitem[Gulyaev et~al.(1986)Gulyaev, Panchuk, Pleshakov, and
  Pyatkes]{gulyaev1986determination}
S.~A. Gulyaev, V.~E. Panchuk, V.~V. Pleshakov, and S.~G. Pyatkes.
\newblock \emph{Astrofiz. Issledovanija, Special Obs.}, \textbf{22}, 3--12
  (1986).

\bibitem[Richardson and Brealey(1973)]{richardson1973jrasc}
E.~H. Richardson and G.~J. Brealey.
\newblock \emph{J. Roy. Astron. Soc. Canada}, \textbf{67}, 165 (1973).

\bibitem[Edvin(1989)]{edvin1989obs}
R.~P. Edvin.
\newblock \emph{The Observatory}, \textbf{109} (1092), 173 (1989).

\bibitem[Panchuk and Klochkova(2013)]{panchuk2013high}
V.~E. Panchuk and V.~G. Klochkova.
\newblock \emph{Bull. of the Crimean Ast. Obs.}, \textbf{109}, 124--135 (2013).

\bibitem[Harrison(1949{\natexlab{b}})]{harrison1949prod2}
G.~R. Harrison.
\newblock \emph{\josa}, \textbf{12}, 6--12 (1949{\natexlab{b}}).

\bibitem[Gerasimov et~al.(1958)Gerasimov, Tel'tevskij, Nesmelov, and
  Sergeev]{gerasimov1958izgot}
F.~M. Gerasimov, I.~A. Tel'tevskij, S.~V. Nesmelov, and V.~P. Sergeev.
\newblock \emph{Opt.-mech. Prom.}, \textbf{12}, 6--12 (1958).

\bibitem[Kopylov and Steshenko(1965)]{kopylov1965ika}
I.~M. Kopylov and N.~V. Steshenko.
\newblock \emph{Izv. Krymskoj Astroph. Obs.}, \textbf{33}, 308 (1965).

\bibitem[Schroeder(1967)]{schroeder1967echelle}
D.~J. Schroeder.
\newblock \emph{\ao}, \textbf{6} (11), 1976--1980 (1967).

\bibitem[Schroeder and Anderson(1971)]{schroeder1971echelle}
D.~J. Schroeder and C.~M. Anderson.
\newblock \emph{\pasp}, \textbf{83} (494), 438 (1971).

\bibitem[McClintock(1979)]{mcclintock1979hires}
W.~McClintock.
\newblock \emph{\pasp}, \textbf{91}, 712--718 (1979).

\bibitem[Hearnshaw(1977)]{hearnshaw1977casseg}
J.~B. Hearnshaw.
\newblock \emph{\pasp}, \textbf{3}, 102 (1977).

\bibitem[McKeith et~al.(1978)McKeith, Dufton, and Kane]{mckeith1978cassegr}
C.~D. McKeith, P.~L. Dufton, and L.~A. Kane.
\newblock \emph{The Observatory}, \textbf{98}, 263--270 (1978).

\bibitem[Latham(1977)]{latham1977inastro}
D.~W. Latham.
\newblock In M.~Duchesne and G.~Lelievre, editors, \emph{Astronomical
  Applications of Image Detectors with Linear Response, Proc. of IAU Coll.},
  volume~40, page~45. Observatoire de Paris-Meudon (1977).

\bibitem[Weiss et~al.(1981)Weiss, Barylak, Hron, and
  Schmiedmayer]{weiss1981pasp}
W.~W. Weiss, M.~Barylak, J.~Hron, and J.~Schmiedmayer.
\newblock \emph{\pasp}, \textbf{93}, 787--794 (1981).

\bibitem[Hunten et~al.(1991)Hunten, Wells, Brown, Schneider, and
  Hilliard]{hunten1991cassegr}
D.~M. Hunten, W.~K. Wells, R.~A. Brown, N.~M. Schneider, and R.~L. Hilliard.
\newblock \emph{\pasp}, \textbf{103}, 1187 (1991).

\bibitem[Argunov et~al.(1967)Argunov, Komarov, and Pozigun]{argunov1967invest}
P.~P. Argunov, N.~S. Komarov, and V.~A. Pozigun.
\newblock \emph{Solar System Research}, \textbf{1}, 93 (1967).

\bibitem[Komarov and Pozigun(1968)]{komarov1968azh}
N.~S. Komarov and N.~S. Pozigun.
\newblock \emph{\azh}, \textbf{45}, 133 (1968).

\bibitem[Mel'nikov and Kuprevich(1956)]{melnikov1956azh}
O.~A. Mel'nikov and N.~F. Kuprevich.
\newblock \emph{\azh}, \textbf{33}, 845 (1956).

\bibitem[Mel'nikov et~al.(1959)Mel'nikov, Kuprevich, and
  Zhukova]{melnikov1959absolute}
O.~A. Mel'nikov, N.~F. Kuprevich, and L.~N. Zhukova.
\newblock \emph{\sovast}, \textbf{3}, 575 (1959).

\bibitem[Jacquinot and Dufour(1948)]{jacquinot1948krcnrs}
P.~Jacquinot and C.~Dufour.
\newblock \emph{J. Rech. Cent. Nat. Rech. Sci.}, \textbf{6}, 91 (1948).

\bibitem[Hiltner and Code(1950)]{hiltner1950josa}
W.~A. Hiltner and A.~D. Code.
\newblock \emph{\josa}, \textbf{40}, 149 (1950).

\bibitem[Geake and Wilcock(1956)]{geake1956mnras}
J.~E. Geake and W.~L. Wilcock.
\newblock \emph{\mnras}, \textbf{116}, 561 (1956).

\bibitem[Boyce et~al.(1973)Boyce, White, Albrecht, and
  Slettebak]{boyce1973pasp}
P.~B. Boyce, N.~M. White, R.~Albrecht, and A.~Slettebak.
\newblock \emph{\pasp}, \textbf{85}, 91 (1973).

\bibitem[Dachs and Schmidt-Kaler(1976)]{dachs1976messenger}
J.~Dachs and T.~Schmidt-Kaler.
\newblock \emph{ESO Messenger}, \textbf{7}, 9--11 (1976).

\bibitem[Liller(1963)]{liller1963ao}
W.~Liller.
\newblock \emph{\ao}, \textbf{2}, 187 (1963).

\bibitem[Namioka(1958)]{namioka1958josa}
T.~Namioka.
\newblock \emph{\josa}, \textbf{49}, 951 (1958).

\bibitem[Kalinenkov and Kharitonov(1967)]{kalinenkov1967afi}
N.~D. Kalinenkov and A.~V. Kharitonov.
\newblock \emph{Trudy AFI AN KazSSR}, \textbf{8}, 128 (1967).

\bibitem[Kharitonov and Klochkova(1972)]{haritonov1972sao}
A.~V. Kharitonov and V.~G. Klochkova.
\newblock \emph{Izv. Spetz. Astrofiz. Observ.}, \textbf{3}, 91 (1972).

\bibitem[Beavers and Eitter(1986)]{beavers1986instr}
W.~I. Beavers and J.~J. Eitter.
\newblock In J.~B. Hearnshaw and P.~L. Cottrell, editors, \emph{Instrumentation
  and research programmes for small telescopes. IAU Symp.}, volume 118,
  page~75. D.~Reidel, Dordrecht (1986).

\bibitem[Gustafsson and Nissen(1972)]{gustafsson1972metal}
B.~Gustafsson and P.~E. Nissen.
\newblock \emph{\aap}, \textbf{19}, 261 (1972).

\bibitem[Nissen(1974)]{nissen1974hetohy}
P.~E. Nissen.
\newblock \emph{\aap}, \textbf{36}, 57--68 (1974).

\bibitem[Nissen(1977)]{nissen1977messenger}
P.~E. Nissen.
\newblock \emph{ESO Messenger}, \textbf{9}, 12--14 (1977).

\bibitem[Oke(1969)]{oke1969pasp}
J.~B. Oke.
\newblock \emph{\pasp}, \textbf{81}, 11 (1969).

\bibitem[Rodgers et~al.(1973)Rodgers, Roberts, Rudge, and
  Stapinski]{rodgers1973pasp}
A.~W. Rodgers, R.~Roberts, P.~T. Rudge, and T.~Stapinski.
\newblock \emph{\pasp}, \textbf{85}, 268 (1973).

\bibitem[Barwig and Schoembs(1986)]{barwig1986instrumentation}
H.~Barwig and R.~Schoembs.
\newblock In J.~B. Hearnshaw and P.~L. Cottrell, editors, \emph{Instrumentation
  and research programmes for small telescopes. IAU Symp.}, volume 118,
  page~61. D.~Reidel, Dordrecht (1986).

\bibitem[Griffin(1969)]{griffin1969photoelectric}
R.~F. Griffin.
\newblock \emph{\mnras}, \textbf{145}, 163 (1969).

\bibitem[Beavers and Eitter(1977)]{beavers1977pasp}
W.~I. Beavers and J.~J. Eitter.
\newblock \emph{\pasp}, \textbf{89}, 733 (1977).

\bibitem[Baranne et~al.(1979)Baranne, Mayor, and Poncet]{baranne1979vistas}
A.~Baranne, M.~Mayor, and J.~L. Poncet.
\newblock \emph{Vistas in Astronomy}, \textbf{23}, 279 (1979).

\bibitem[Baranne et~al.(1977)Baranne, Mayor, and Poncet]{baranne1977sur}
A.~Baranne, M.~Mayor, and J.~L. Poncet.
\newblock \emph{Acad. des Sci. (Paris), Comptes Rendus, Serie B~-- Sci.
  Physiq.}, \textbf{285} (4), 117--120 (1977).

\bibitem[Imbert and Pr\'evot(1981)]{imbert1981first}
M.~Imbert and L.~Pr\'evot.
\newblock \emph{ESO Messenger}, \textbf{25}, 6 (1981).

\bibitem[Beavers et~al.(1980)Beavers, Eitter, Carr, and Cook]{beavers1980aj}
W.~I. Beavers, J.~J. Eitter, P.~H. Carr, and B.~C. Cook.
\newblock \emph{\apj}, \textbf{238}, 349 (1980).

\bibitem[Tokovinin(1987)]{tokovinin31stellar}
A.~A. Tokovinin.
\newblock \emph{\sovast}, \textbf{31} (1), 98 (1987).

\bibitem[Sloane et~al.(1969)Sloane, Fine, Phillips, and Harvit]{sloane1969ao}
N.~J.~A. Sloane, T.~Fine, P.~G. Phillips, and M.~O. Harvit.
\newblock \emph{\ao}, \textbf{8}, 2103 (1969).

\bibitem[Harwit(1971)]{harwit1971ao}
M.~Harwit.
\newblock \emph{\ao}, \textbf{10}, 1415 (1971).

\bibitem[{Ming~Hing~Tai} et~al.(1975){Ming~Hing~Tai}, Briotta~Jr., Kamath, and
  Harwit]{minghingtai1975ao}
{Ming~Hing~Tai}, D.~A. Briotta~Jr., N.~S. Kamath, and M.~Harwit.
\newblock \emph{\ao}, \textbf{14}, 2533 (1975).

\bibitem[Wood(1968)]{wood1968pasp}
H.~J. Wood.
\newblock \emph{\pasp}, \textbf{80} (477), 647 (1968).

\bibitem[Esipov(1963)]{esipov1963novtech}
V.~F. Esipov.
\newblock In \emph{Novaya tehnika v astronomii}, volume~1, page 165. Leningrad:
  Nauka (1963).

\bibitem[Lallemand and Duchesne(1956)]{lallemand1956astronomie}
A.~Lallemand and M.~Duchesne.
\newblock \emph{Publ. de l'Observatoire de Haute-Provence}, \textbf{3} (50), 3
  (1956).

\bibitem[Baranne and Duchesne(1972)]{baranne1972auxinstr}
A.~Baranne and M.~Duchesne.
\newblock In S.~Lautsen and A.~Reiz, editors, \emph{Auxiliary Instrumentation
  for Large Telescopes. Proc. ESO/CERN Conf.}, page 241. Geneva, Switzerland
  (1972).

\bibitem[Breysacher(1976)]{breysacher1976messenger}
J.~Breysacher.
\newblock \emph{ESO Messenger}, \textbf{5}, 3 (1976).

\bibitem[Baranne et~al.(1974)Baranne, Carozzi, Comte, Courtes, Deharveng,
  Duflot, Monnet, and Pellet]{baranne1974preliminary}
A.~Baranne, N.~Carozzi, G.~Comte, G.~Courtes, J.~M. Deharveng, R.~Duflot,
  G.~Monnet, and A.~Pellet.
\newblock In A.~Reiz, editor, \emph{Research Programmes for the New Large
  Telescopes, Proceedings of the ESO/SRC/CERN Conference}, page 231. Geneva,
  Switzerland (1974).

\bibitem[Margon et~al.(1979)Margon, Ford, Katz, Kwitter, Ulrich, Stone, and
  Klemola]{margon1979aj}
B.~Margon, H.~C. Ford, J.~I. Katz, K.~B. Kwitter, R.~G. Ulrich, R.~P.~S. Stone,
  and A.~Klemola.
\newblock \emph{\aj}, \textbf{230}, 41 (1979).

\bibitem[McNall et~al.(1972)McNall, Michalski, and Miedaner]{mcnall1972pasp}
J.~F. McNall, D.~E. Michalski, and T.~L. Miedaner.
\newblock \emph{\pasp}, \textbf{84} (497), 145 (1972).

\bibitem[Torres-Dodgen and Weaver(1993)]{torresdodgen1993atlas}
A.~V. Torres-Dodgen and W.~B. Weaver.
\newblock \emph{\pasp}, \textbf{105}, 693 (1993).

\bibitem[Schectman and Hiltner(1976)]{schectman1976pasp}
S.~A. Schectman and W.~A. Hiltner.
\newblock \emph{\pasp}, \textbf{88}, 960 (1976).

\bibitem[Schectman(1984)]{schectman1984spie}
S.~Schectman.
\newblock \emph{Proc. SPIE}, \textbf{445}, 128 (1984).

\bibitem[Mochnacki et~al.(1986)Mochnacki, Chew, Kunowski, Hawker, Kamper,
  Blyth, Zerafa, and Platzer]{mochnacki1985ddo}
S.~W. Mochnacki, S.~Chew, W.~Kunowski, F.~Hawker, K.~Kamper, D.~Blyth,
  L.~Zerafa, and A.~Platzer.
\newblock In J.~B. Hearnshaw and P.~L. Cottrell, editors, \emph{Proceedings of
  the 118th. Symposium of the International Astronomical Union}, page 461.
  D.~Reidel, Dordrecht, Holland (1986).

\bibitem[Gray(1976)]{gray1976theobs}
D.~F. Gray.
\newblock \emph{The Observation and Analysis of Stellar Photospheres}.
\newblock Univ. of Western Ontario. John Wiley and Sons Inc. (1976).

\bibitem[Gray(1978)]{gray1978high}
D.~F. Gray.
\newblock In M.~Hack, editor, \emph{High Resolution Spectrometry, Proc. of the
  4-th Colloq. on Astrophys}, page 268. Osservatorio Astronomico di Trieste
  (1978).

\bibitem[Gray(1986)]{gray1986instrum}
D.~F. Gray.
\newblock In J.~B. Hearnshaw and P.~L. Cottrell, editors, \emph{Instrumentation
  and research programmes for small telescopes, IAU Symp.}, volume 118, page
  401. D.Reidel, Dordrecht (1986).

\bibitem[Enard(1979)]{enard1979messenger}
D.~Enard.
\newblock \emph{ESO Messenger}, \textbf{17}, 32--33 (1979).

\bibitem[Andersen(1977)]{andersen1977messenger}
T.~Andersen.
\newblock \emph{ESO Messenger}, \textbf{10}, 21--23 (1977).

\bibitem[Andersen(1979)]{andersen1979messenger}
T.~Andersen.
\newblock \emph{ESO Messenger}, \textbf{16}, 37--38 (1979).

\bibitem[Furenlid(1984)]{furenlid1984pasp}
I.~Furenlid.
\newblock \emph{\pasp}, \textbf{96}, 325 (1984).

\bibitem[McDavid(1986)]{mcdavid1986instr}
D.~A. McDavid.
\newblock In J.~B. Hearnshaw and P.~L. Cottrell, editors, \emph{Instrumentation
  and research programmes for small telescopes, IAU Symp.}, volume 118, page
  457. D. Reidel, Dordrecht (1986).

\bibitem[Denby et~al.(1986)Denby, Dalglish, Meadows, and
  Taylor]{denby1986instr}
B.~Denby, R.~Dalglish, V.~Meadows, and K.~N.~R. Taylor.
\newblock In J.~B. Hearnshaw and P.~L. Cottrell, editors, \emph{Instrumentation
  and research programmes for small telescopes, IAU Symph}, volume 118, page
  439. D.~Reidel, Dordrecht (1986).

\bibitem[Rakos et~al.(1990)Rakos, Weiss, Muller, Pressberger, Wachtler,
  Schombert, and Kreidl]{rakos1990pasp}
K.~D. Rakos, W.~W. Weiss, S.~Muller, R.~Pressberger, P.~Wachtler, J.~M.
  Schombert, and T.~J. Kreidl.
\newblock \emph{\pasp}, \textbf{102}, 674 (1990).

\bibitem[Fabricant et~al.(1998)Fabricant, Cheimets, Caldwell, and
  Geary]{fabrikant1998pasp}
D.~Fabricant, P.~Cheimets, N.~Caldwell, and J.~Geary.
\newblock \emph{\pasp}, \textbf{110}, 79 (1998).

\bibitem[Latham(1982)]{latham1982instr}
D.~Latham.
\newblock In C.~M. Humphries, editor, \emph{Instrumentation for Astronomy with
  Large Telescopes, IAU Coll.}, page 259. Dordrecht: Reidel (1982).

\bibitem[Munari and Lattanzi(1992)]{munari1992pasp}
U.~Munari and M.~G. Lattanzi.
\newblock \emph{\pasp}, \textbf{104}, 121 (1992).

\bibitem[Munari and Valisa(2014)]{munari2014multimode}
U.~Munari and P.~Valisa.
\newblock \emph{Contr. of the Astron. Obs. Skalnat\'e Pleso}, \textbf{43} (3),
  174--182 (2014).

\bibitem[Panchuk et~al.(2015{\natexlab{a}})Panchuk, Klochkova, Sachkov, Verich,
  and Yushkin]{panchuk2015podvesnoj}
V.~E. Panchuk, V.~G. Klochkova, M.~E. Sachkov, Y.~B. Verich, and M.~V. Yushkin.
\newblock In \emph{The present and future of small and medium-size telescopes},
  page~70. Nizhny Arkhyz (2015{\natexlab{a}}).

\bibitem[Panchuk et~al.(2017{\natexlab{a}})Panchuk, Klochkova, and
  Yushkin]{panchuk2017hires}
V.~E. Panchuk, V.~G. Klochkova, and M.~V. Yushkin.
\newblock \emph{Astronomy Reports}, \textbf{61} (9), 820--831
  (2017{\natexlab{a}}).

\bibitem[Afanasiev et~al.(2016)Afanasiev, Dodonov, Amirkhanyan, and
  Moiseev]{afanasiev2016adam}
V.~L. Afanasiev, S.~N. Dodonov, V.~R. Amirkhanyan, and A.~V. Moiseev.
\newblock \emph{\ab}, \textbf{71} (4), 479--488 (2016).

\bibitem[Angel and Gresham(1979)]{angel1979operation}
J.~R.~P. Angel and M.~S. Gresham.
\newblock \emph{\apj Part~1}, \textbf{229}, 1074--1078 (1979).

\bibitem[Angel et~al.(1977)Angel, Adams, Boroson, and Moore]{angel1977very}
J.~R.~P. Angel, M.~T. Adams, T.~A. Boroson, and R.~L. Moore.
\newblock \emph{\apj Part~1}, \textbf{218}, 776--782 (1977).

\bibitem[Furenlid and Cardona(1988)]{furenlid1988pasp}
I.~Furenlid and O.~Cardona.
\newblock \emph{\pasp}, \textbf{100}, 1001 (1988).

\bibitem[Barry et~al.(2002)Barry, Bagnuolo~Jr., and Riddle]{barry2002pasp}
D.~J. Barry, W.~G. Bagnuolo~Jr., and R.~L. Riddle.
\newblock \emph{\pasp}, \textbf{114}, 198 (2002).

\bibitem[Bagnuolo et~al.(1990)Bagnuolo, Furenlid, Gies, and
  Barry]{bagnuolo1990pasp}
W.~G. Bagnuolo, I.~Furenlid, D.~R. Gies, and D.~J. Barry.
\newblock \emph{\pasp}, \textbf{102}, 604 (1990).

\bibitem[Kershaw and Hearnshaw(1989)]{kershaw1989southern}
G.~M. Kershaw and J.~B. Hearnshaw.
\newblock \emph{Southern Stars}, \textbf{33}, 89 (1989).

\bibitem[Murdoch et~al.(1993)Murdoch, Hearnshaw, and Clark]{murdoch1993search}
K.~A. Murdoch, J.~B. Hearnshaw, and M.~Clark.
\newblock \emph{\aj}, \textbf{413}, 349 (1993).

\bibitem[Hearnshaw et~al.(2002)Hearnshaw, Barnes, Kershaw,
  et~al.]{hearnshaw2002exast}
J.~B. Hearnshaw, S.~I. Barnes, G.~M. Kershaw, et~al.
\newblock \emph{Experimental Astron}, \textbf{13}, 59 (2002).

\bibitem[Mandel(1988)]{mandel1988impact}
H.~Mandel.
\newblock In G.~{Cayrel de Strobel} and M.~Spite, editors, \emph{Proc. of the
  IAU Symp.}, volume 132, page~9. Springer Netherlands (1988).

\bibitem[Wolf et~al.(1993)Wolf, Mandel, Stahl, Kaufer, Szeifert, Gang,
  Gummersbach, and Kovacs]{wolf1993messenger}
B.~Wolf, H.~Mandel, O.~Stahl, A.~Kaufer, T.~Szeifert, T.~Gang, C.~A.
  Gummersbach, and J.~Kovacs.
\newblock \emph{ESO Messenger}, \textbf{74}, 19 (1993).

\bibitem[Lub(1979)]{lub1979messenger}
J.~Lub.
\newblock \emph{The Messenger}, \textbf{19}, 1--4 (1979).

\bibitem[Slechta and Skoda(2002)]{slechta2002pasi}
M.~Slechta and P.~Skoda.
\newblock \emph{Publ. Astron. Inst. ASCR}, \textbf{90}, 1 (2002).

\bibitem[Vanzi et~al.(2012)Vanzi, Chacon, Helminiak, Baffico, Rivinius, Stefl,
  Baade, Avila, and Guirao]{vanzi2012pucheros}
L.~Vanzi, J.~Chacon, K.~G. Helminiak, M.~Baffico, T.~Rivinius, S.~Stefl,
  D.~Baade, G.~Avila, and C.~Guirao.
\newblock \emph{\mnras}, \textbf{424}, 2770--2777 (2012).

\bibitem[Baudrand and Bohm(1992)]{baudrand1992aap}
J.~Baudrand and T.~Bohm.
\newblock \emph{\aap}, \textbf{259}, 711 (1992).

\bibitem[Morrison(1995)]{morrison1995bull}
N.~D. Morrison.
\newblock \emph{Bull. of the Astron. Soc.}, \textbf{27} (1), 647--653 (1995).

\bibitem[Marino et~al.(1998)Marino, Catalano, Frasca, and
  E.Marilli]{marino1998ibvs}
G.~Marino, S.~Catalano, A.~Frasca, and E.Marilli.
\newblock \emph{Inf. Bull. on Var. Stars}, \textbf{4599}, 1--3 (1998).

\bibitem[Queloz et~al.(2000)Queloz, Mayor, Weber, et~al.]{queloz2000aa}
D.~Queloz, M.~Mayor, L.~Weber, et~al.
\newblock \emph{\aap}, \textbf{354}, 99--102 (2000).

\bibitem[Baranne et~al.(1996)Baranne, Queloz, Mayor, Adrianzyk, Knispel,
  Kohler, Lacroix, Meunier, Rimbaud, and Vin]{baranne1996aass}
A.~Baranne, D.~Queloz, M.~Mayor, G.~Adrianzyk, G.~Knispel, D.~Kohler,
  D.~Lacroix, J.-P. Meunier, G.~Rimbaud, and A.~Vin.
\newblock \emph{\aas}, \textbf{119}, 373 (1996).

\bibitem[Weber et~al.(2000)Weber, Blecha, Davignon, Maire, Queloz, Russiniello,
  and Simond]{weber2000fully}
L.~Weber, A.~Blecha, G.~Davignon, C.~Maire, D.~Queloz, G.~B. Russiniello, and
  G.~Simond.
\newblock \emph{Proc. SPIE}, \textbf{4009}, 61--70 (2000).

\bibitem[Strassmeier et~al.(2001)Strassmeier, Granzer, Weber, Woche,
  et~al.]{strassmeier2001stella}
K.~G. Strassmeier, T.~Granzer, M.~Weber, M.~Woche, et~al.
\newblock \emph{\an}, \textbf{322} (5/6), 287--294 (2001).

\bibitem[Strassmeier et~al.(2004)Strassmeier, Granzer, Weber,
  et~al.]{strassmeier2004stella}
K.~G. Strassmeier, T.~Granzer, M.~Weber, et~al.
\newblock \emph{\an}, \textbf{325} (6/8), 527--532 (2004).

\bibitem[Raskin and Winkel(2008)]{raskin2008spie}
G.~Raskin and H.~V. Winkel.
\newblock \emph{Proc SPIE}, \textbf{7014}, 178 (2008).

\bibitem[Panchuk et~al.(2011)Panchuk, Yushkin, and Yakopov]{panchuk2011high}
V.~E. Panchuk, M.~V. Yushkin, and M.~V. Yakopov.
\newblock \emph{\ab}, \textbf{66} (3), 355--370 (2011).

\bibitem[Queloz(1995)]{queloz1995new}
D.~Queloz.
\newblock In A.~G.~D. Philip, K.~A. Janes, and A.~R. Upgren, editors, \emph{New
  Developments in Array Technology and Applications}, page 221. IAU,
  Netherlands (1995).

\bibitem[Panchuk et~al.(2015{\natexlab{b}})Panchuk, Yushkin, Klochkova,
  Yakopov, and Verich]{panchuk2015design}
V.~E. Panchuk, M.~V. Yushkin, V.~G. Klochkova, G.~V. Yakopov, and Y.~B. Verich.
\newblock \emph{\ab}, \textbf{70} (2), 226--231 (2015{\natexlab{b}}).

\bibitem[Punanova and Krushinsky(2013)]{punanova2013ika}
A.~F. Punanova and V.~V. Krushinsky.
\newblock \emph{Izv. Krymskoj AO}, \textbf{109} (2), 51--53 (2013).

\bibitem[Gibson et~al.(2012)Gibson, Barnes, Hearnshaw, Nield, Cochrane, and
  Grobler]{gibson2012kiwispec}
S.~Gibson, S.~I. Barnes, J.~Hearnshaw, K.~Nield, D.~Cochrane, and D.~Grobler.
\newblock \emph{Proc. SPIE}, \textbf{8446}, 844648 (2012).

\bibitem[Barnes et~al.(2012)Barnes, Gibson, Nield, and
  Cochrane]{barnes2012kiwispec}
S.~I. Barnes, S.~Gibson, K.~Nield, and D.~Cochrane.
\newblock In \emph{Ground-based and Airborne Instrumentation for Astronomy IV},
  volume 8446, page 844688. Proc. SPIE (2012).

\bibitem[Panchuk et~al.(2015{\natexlab{c}})Panchuk, Klochkova, Sachkov, and
  Yushkin]{panchuk2015doppler}
V.~E. Panchuk, V.~G. Klochkova, M.~E. Sachkov, and M.~V. Yushkin.
\newblock \emph{Solar System Research}, \textbf{49} (6), 420--429
  (2015{\natexlab{c}}).

\bibitem[Baudrand and Walker(2001)]{baudrant2001modal}
J.~Baudrand and G.~A.~H. Walker.
\newblock \emph{\pasp}, \textbf{113} (785), 851--858 (2001).

\bibitem[Schwab et~al.(2012)Schwab, Leon-Saval, Betters, Bland-Hawthorn, and
  Mahadevan]{schwab2012single}
C.~Schwab, S.~G. Leon-Saval, C.~H. Betters, J.~Bland-Hawthorn, and
  S.~Mahadevan.
\newblock \emph{Proc. IAU Symposium}, \textbf{293} (2012).

\bibitem[Klochkova et~al.(2005)Klochkova, Panchuk, Romanenko, and
  Najdenov]{klochkova2005bull}
V.~G. Klochkova, V.~E. Panchuk, V.~P. Romanenko, and I.~D. Najdenov.
\newblock \emph{Bull. Spec. Astroph. Obs.}, \textbf{58}, 132--144 (2005).

\bibitem[Kawabata et~al.(1999)Kawabata, Okasaki, Akitava, Hirakata, Hirata,
  Ikeda, Kondon, Masuda, and Seki]{kawabata1999pasp}
K.~S. Kawabata, A.~Okasaki, H.~Akitava, N.~Hirakata, R.~Hirata, Y.~Ikeda,
  M.~Kondon, S.~Masuda, and M.~Seki.
\newblock \emph{\pasp}, \textbf{111}, 898 (1999).

\bibitem[Serkowski et~al.(1975)Serkowski, Mathewson, and
  Ford]{serkovski1975apj}
K.~Serkowski, D.~S. Mathewson, and V.~Ford.
\newblock \emph{\apj}, \textbf{196}, 261 (1975).

\bibitem[Panchuk et~al.(2017{\natexlab{b}})Panchuk, Klochkova, Yushkin,
  Yakopov, and Verich]{panchuk2017izvvuz}
V.~E. Panchuk, V.~G. Klochkova, M.~V. Yushkin, G.~V. Yakopov, and Y.~B. Verich.
\newblock \emph{Izv. Vyshih Uchebnyh Zavedenij. Priborostroenie}, \textbf{60}
  (1), 53--62 (2017{\natexlab{b}}).

\bibitem[Davidson et~al.(2014)Davidson, Bjorkman, Hoffman, Bjorkman,
  et~al.]{davidson2014hpol}
J.~W. Davidson, K.~S. Bjorkman, J.~L. Hoffman, J.~E. Bjorkman, et~al.
\newblock \emph{Journal of Astron. Instr.}, \textbf{3} (3n04), 1450009 (2014).

\bibitem[Fellgett(1958)]{fellgett1958jpr}
P.~Fellgett.
\newblock \emph{Journal de Physique et le Radium}, \textbf{19}, 187--191
  (1958).

\bibitem[Jacquinot(1957)]{jacquinot1957josa}
P.~Jacquinot.
\newblock \emph{\josa}, \textbf{44}, 761 (1957).

\bibitem[Geake and Wilcock(1957)]{geake1957mnras}
J.~E. Geake and W.~L. Wilcock.
\newblock \emph{\mnras}, \textbf{117}, 380--383 (1957).

\bibitem[Tarasov(1968)]{tarasov1968spectr}
K.~N. Tarasov.
\newblock \emph{Spectral'nye pribory}.
\newblock Leningrad: Mashinostroenie (1968).

\bibitem[Mertz(1958)]{mertz1958jpl}
L.~Mertz.
\newblock \emph{Journal de Physique et le Radium}, \textbf{19} (3), 233 (1958).

\bibitem[Serkowski(1972)]{serkowski1972pasp}
K.~Serkowski.
\newblock \emph{\pasp}, \textbf{84}, 649 (1972).

\bibitem[Serkowski(1978)]{serkowski1978high}
K.~Serkowski.
\newblock In M.~Hack, editor, \emph{High Resolution Spectrometry, Proc. of the
  4th Colloq. on Astrophys.}, page 245. Osservatorio Astronomico di Trieste
  (1978).

\bibitem[McMillan et~al.(1993)McMillan, Moore, Perry, and
  Smith]{mcmillan1993aj}
R.~S. McMillan, T.~Moore, M.~L. Perry, and P.~H. Smith.
\newblock \emph{\aj}, \textbf{403}, 801 (1993).

\bibitem[Panchuk(Spec. Astrophys. Obs, Nizhny Arkhyz,
  2000)]{panchuk2000prepsao144}
V.~E. Panchuk.
\newblock Preprint No.~144, SAO RAS (Spec. Astrophys. Obs, Nizhny Arkhyz,
  2000).

\bibitem[Erskine(2003)]{erskine2003pasp}
D.~J. Erskine.
\newblock \emph{\pasp}, \textbf{115}, 255 (2003).

\bibitem[Frandsen et~al.(1993)Frandsen, Douglas, and Butcher]{frandsen1993aap}
S.~Frandsen, N.~Douglas, and H.~Butcher.
\newblock \emph{\aap}, \textbf{279}, 310 (1993).

\bibitem[Ge et~al.(2006)Ge, van Eyken, Mahadevan, et~al.]{jian2006first}
J.~Ge, J.~van Eyken, S.~Mahadevan, et~al.
\newblock \emph{\apj}, \textbf{648} (1), 683--695 (2006).

\bibitem[Castley(1972)]{castley1972pasa}
J.~C. Castley.
\newblock \emph{Publ. Astron. Soc. Australia}, \textbf{2}, 137 (1972).

\bibitem[Castley and Watson(1980)]{castley1980aas}
J.~C. Castley and R.~D. Watson.
\newblock \emph{\aas}, \textbf{41}, 397 (1980).

\bibitem[Gull et~al.(1974)Gull, O'Dell, and Parker]{gull1974icarus}
T.~R. Gull, C.~R. O'Dell, and R.~A.~R. Parker.
\newblock \emph{Icarus}, \textbf{21} (3), 213 (1974).

\bibitem[Miroshnichenko et~al.(2013)Miroshnichenko, Pasechnik, Manset,
  et~al.]{miroshnichenko2013periastron}
A.~S. Miroshnichenko, A.~V. Pasechnik, N.~Manset, et~al.
\newblock \emph{\apj}, \textbf{766} (2), 119 (2013).

\bibitem[Avila et~al.(2007)Avila, Burwitz, Guirao, Rodriguez, Shida, and
  Baade]{avila2007high}
G.~Avila, V.~Burwitz, C.~Guirao, J.~Rodriguez, R.~Shida, and D.~Baade.
\newblock \emph{ESO Messenger}, \textbf{129}, 62 (2007).

\bibitem[Koz\l{}owski et~al.(2014)Koz\l{}owski, Konacki, Ratajczak, Sybilski,
  Paw\l{}aszek, and He\l{}miniak]{kozlowski2014baches}
S.~K. Koz\l{}owski, M.~Konacki, M.~Ratajczak, P.~Sybilski, R.~K. Paw\l{}aszek,
  and K.~G. He\l{}miniak.
\newblock \emph{\mnras}, \textbf{443} (1), 158--167 (2014).

\bibitem[Panchuk et~al.(2015{\natexlab{d}})Panchuk, Klochkova, and
  Marchenko]{panchuk2015spectrograph}
V.~E. Panchuk, V.~V. Klochkova, V.~G.~Komarov, and D.~V. Marchenko.
\newblock In \emph{The present and future of small and medium-size telescopes},
  page~69. Nizhny Arkhyz (2015{\natexlab{d}}).

\bibitem[Panchuk and Verich(2015)]{panchuk2015autocoll}
V.~E. Panchuk and Y.~B. Verich.
\newblock In \emph{The present and future of small and medium-size telescopes},
  pages 66--67. Nizhny Arkhyz (2015).

\bibitem[Panchuk et~al.(2014)Panchuk, Chuntonov, and Naidenov]{panchuk2014main}
V.~E. Panchuk, G.~A. Chuntonov, and I.~D. Naidenov.
\newblock \emph{\ab}, \textbf{69} (3), 339--355 (2014).

\bibitem[Thizy(2011)]{thizy2011spectrographs}
F.~Thizy, O.;~Cochard.
\newblock In \emph{Active OB stars\ldots, Proc. of the IAU Symp.}, volume 272,
  pages 282--283. Cambridge University Press (2011).

\bibitem[Kaye et~al.(2006)Kaye, Vanaverbeke, and Innis]{kaye2006high}
T.~G. Kaye, S.~Vanaverbeke, and J.~Innis.
\newblock \emph{J. Br. Astron. Assoc.}, \textbf{116} (2), 78--83 (2006).

\bibitem[Panchuk et~al.(2020)Panchuk, {{Yu.~Yu. Balega}}, Klochkova, and
  Sachkov]{panchuk2020issledovanie}
V.~E. Panchuk, {{Yu.~Yu. Balega}}, V.~G. Klochkova, and M.~E. Sachkov.
\newblock \emph{Uspehi Fizicheskih Nauk}, \textbf{190} (6), 605--626 (2020).
\newblock \doi{10.3367/UFNr.2019.07.038597}.
\newblock URL \url{https://ufn.ru/ru/articles/2020/6/c/}.

\bibitem[Perruchot et~al.(2008)Perruchot, Kohler, Bouchy,
  et~al.]{perruchot2008sophie}
S.~Perruchot, D.~Kohler, F.~Bouchy, et~al.
\newblock In I.~S. McLean and M.~M. Casali, editors, \emph{Ground-based and
  Airborne Instrumentation for Astronomy II}, volume 7014, page 7014J. Proc.
  SPIE (2008).

\bibitem[Bouchy et~al.(1999)Bouchy, Connes, and Bertaux]{bouchy1999iaucoll}
F.~Bouchy, P.~Connes, and J.~L. Bertaux.
\newblock In J.~B. Hearnshaw and C.~D. Scarfe, editors, \emph{Precise Stellar
  Radial Velocities, ASP Conf. Ser. No\,185}, volume 170, page~22. IAU Coll.
  (1999).

\bibitem[Nagulin et~al.(1980)Nagulin, Pavlyucheva, and
  Yakovlev]{nagulin1980echelle}
{\relax Yu.~S}.~Nagulin, N.~K. Pavlyucheva, and E.~A. Yakovlev.
\newblock \emph{Optika i Spectroskopiya}, \textbf{49} (5), 987--989 (1980).

\bibitem[Gerasimov et~al.(1970)Gerasimov, Yakovlev, Pejsahson, and
  Koshelev]{gerasimov1970ois}
F.~M. Gerasimov, E.~A. Yakovlev, I.~V. Pejsahson, and B.~V. Koshelev.
\newblock \emph{Optika i Spectroskopiya}, \textbf{28}, 790 (1970).

\bibitem[Panchuk et~al.(2018)Panchuk, Yakopov, Klochkova, and
  Yushkin]{panchuk2018kalibrovka}
V.~E. Panchuk, M.~V. Yakopov, V.~G. Klochkova, and M.~V. Yushkin.
\newblock \emph{Pribory i tehnika eksperimenta}, \textbf{4}, 106--110 (2018).

\end{thebibliography}

\end{document}